\documentclass[twocolumn,aps,pre,superscriptaddress,floatfix]{revtex4-1}
\usepackage{graphicx}
\usepackage{epsfig}
\usepackage{thumbpdf}
\usepackage{amsfonts}
\usepackage{times}     
\usepackage{float}
\usepackage{indentfirst}   %
\usepackage{amsmath}
\usepackage{epstopdf}
\usepackage{textcomp}
\usepackage{tocvsec2}
\usepackage{epsfig}
\usepackage{xcolor}
\usepackage{url}
\usepackage{comment}
\usepackage{CJK} 

\usepackage[section]{placeins}
\makeatletter\AtBeginDocument{%
     \expandafter\renewcommand\expandafter\subsection\expandafter
       {\expandafter\@fb@subsecFB\subsection}%
     \newcommand\@fb@subsecFB{\FloatBarrier
     \gdef\@fb@afterHHook{\@fb@topbarrier \gdef\@fb@afterHHook{}}}
     \g@addto@macro\@afterheading{\@fb@afterHHook}
     \gdef\@fb@afterHHook{}
  }
\usepackage[colorlinks=true,linkcolor=black]{hyperref} 

\newcommand{\be}{\begin{equation}}
\newcommand{\ee}{\end{equation}}

\begin{document}
\begin{CJK}{UTF8}{gbsn}

\title{Cooperator-driven and defector-driven punishments: How do they influence cooperation?}

\author{Pengbi Cui (崔鹏碧)}\email{cuisir610@gmail.com}
\affiliation{School of Astronautics, Northwestern Polytechnical University, Xi'an, 710072, PR China}
\affiliation{School of Computer Science and Engineering, University of Electronic Science and Technology of China, Chengdu 610054, China}

\author{Zhi-Xi Wu}
\affiliation{Institute of Computational Physics and Complex Systems, Lanzhou University, Lanzhou Gansu 730000, China and Key Laboratory for Magnetism and Magnetic Materials of the Ministry of Education, Lanzhou University, Lanzhou Gansu 730000, China}

\author{Tao Zhou}
\affiliation{School of Computer Science and Engineering, University of Electronic Science and Technology of China, Chengdu 610054, China}
\affiliation{Big Data Research Center, University of Electronic Science and Technology of China, Chengdu 611731, China}
\affiliation{Institute of Fundamental and Frontier Science, University of Electronic Science and Technology of China, Chengdu 610073, China}

\author{Xiaojie Chen}
\affiliation{School of Mathematical Sciences, University of Electronic Science and Technology of China, Chengdu, 611731, China}

\begin{abstract}
Economic studies have shown that there are two types of regulation schemes which can be considered as a vital part of today's global economy: self-regulation enforced by self-regulation organizations to govern industry practices, and government regulation which is considered as another scheme to sustain corporate adherence. An outstanding problem of particular interest is to understand quantitatively the role of these regulation schemes in evolutionary dynamics. Typically, punishment usually occurs for enforcement of regulations. Taking into account both types of punishments to curve the regulations, we develop a game model where six evolutionary situations with corresponding combinations of strategies are considered. Furthermore, a semi-analytical method is developed to allow us to give an accurate estimations of the boundaries between the phases of full defection and nondefection. We find that, associated with the evolutionary dynamics, for infinite well-mixed population, the mix of both punishments performs better than one punishment alone in promoting public cooperation, but for networked population the cooperator-driven punishment turns out to be a better choice. We also uncover monotonous facilitating effects of synergy effect, punishment fine and feedback sensitivity on the public cooperation for infinite well-mixed population. Conversely, for networked population an optimal intermediate range of feedback sensitivity is needed to best promote punishers' populations. Overall, networked structure is overall more favorable for punishers and further for public cooperation, because of both network reciprocity and mutualism between punishers and cooperators who do not punish defectors. We provide physical understandings of the observed phenomena, through a detailed statistical analysis of frequencies of different strategies and spatial pattern formations in different evolution situations. These results provide valuable insights into how to select and enforce suitable regulation measures to let public cooperation keep prevalent, which has potential implications not only to self-regulation, but also to other topics in economics and social science.
\end{abstract}

\maketitle
\end{CJK}

\section{Introduction}
\label{sec:introduction}
In the field of evolutionary game theory, the conventional social dilemma i.~e. the first-order social dilemma means that the well-being of the population depend only on the level of cooperation while defection is the best choice for one individual. Besides the mechanisms to sustain or promote cooperative behaviors such as kin selection~\cite{foster2006}, reputation~\cite{alexander1987}, group selection, reciprocity~\cite{axelrod1984,nowak2006}, punishment has also been widely approved as an available rule to alleviate this public good problem~\cite{fehr2000,osborne2004}. Large number of related studies have been proceeded to center on how punishment rules govern the evolution of the game systems~\cite{fehr2000,fehr2002,gachter2008,herrmann2008}. At the same time, these studies have affirmed that punishment is a useful tool to repel defection behaviors and to facilitate cooperation of the population, through both empirical experiments and theoretical analysis. However, second-order free-riding (i.e., second-order dilemma) arising from the fact that punishers have to bear extra substantial punishment cost is a non-ignorable impediment to the evolutionary stability of punishment. Since this would weaken punishers' persistent monitoring ability and sanctions on wrong-doers~\cite{panchanathan2004,Sigmund2010,perc2012}. Aiming to address this issue, some researchers have tried to seek more effective specific strategies or mechanisms~\cite{hauert2002,mathew2009,Perc2010,boyd2010,cui2014,chen2014}. 

An issue of growing interest in current research of game theory is that humans prefer pool punishment over peer punishment for maintaining the commons~\cite{traulsen2012}. Unlike peer punishment in which the punishment act is carried out by peers, measures of pool punishment are usually outsourced and carried out by a paid organization which collects punishment costs (i.e., taxes) from the cooperators who are willing to eliminate the defectors from the population~\cite{Sigmund2010,Szolnoki2011}. It is convinced that these cooperators can be regarded as punishers to some extent, who can commonly share the cost of pool punishment. Within this game framework, in some cases  consideration of punishment strategies can solve the 'second-order free-rider problem' in the presence of a segregation of behavioral strategies~\cite{helbing2010}, or if punishment fine is large enough~\cite{helbing2010p}. Also, it has been found that prosocial punishers can outperform combination of positive and negative reciprocity, which while is able to invade defectors~\cite{szolnoki2013}. Especially, recent studies highlight 'adaptive punishment' is good at facilitating public cooperation or even double-edged, where the punishers condition their sanctioning activities against antisocial behaviors on one threshold relating to the success or abundance of cooperators/defectors within the groups~\cite{perc2012self,Huang2018,wu2018}. In reality pool punishment is widely exploited by many authorities to mitigate the free riders' destructive potential; regardless of the punishment being direct, indirect, first order or second order~\cite{yamagishi1986,o2009,Sigmund2010,sasaki2012}. The cost of pool punishment is commonly shared, which thus could reduce both financial burdens and the risk of being revenged as much as possible~\cite{nikiforakis2008}. Implement institutions of the pool punishment are also easily to be established to ensure the fairness so that defectors are timely identified and punished. As operation institutions of pool punishment, third organizations such as modern courts, the police system and regulation organizations are developed to carry out punitive measures, so as to alleviate the problems of ''second-order free-riders'', antisocial punishment~\cite{herrmann2008,rand2011} and retaliation~\cite{nikiforakis2008}. Therefore, pool punishment has widely gained more attentions than before as an important symbol of modern civilized society.

Moreover, as an important background to motivate our present study, there are one realistic case which must be stated with respect to punishment measures whose executions are strongly dependent on the abundance of free-riding behaviors in the systems. 

In commerce self-regulation is an important mechanism for governing industry practices, owning many benefits over government regulation for consumers, businesses, the government, and the economy as a whole~\cite{Short2010}. The incentive for the private sector to undertake self-regulatory actions i.e., to develop and comply with standards is that they are incentivized by customer expectations and threatened by possible government regulation, critical public opinion, which actually issues statements of concern for public welfare~\cite{Sharma2010}. Further, self regulation is the process whereby an organization or company is enforced by self-regulation organization (SRO) to monitor its own adherence to legal, ethical or safety standards, rather than a third-party and independent agency such as a governmental entity monitor to enforce those standards~\cite{wikisr}. Self-regulation is a win-win for both businesses, consumers and government. Businesses benefit from not only regulations that are predictable and reasonable, as opposed to command and control rules that are often burdensome and expensive to comply with; but also more efficient enforcement approaches which allow them to better manage their scarce resources~\cite{Sharma2010,Dashwood2014} and to increase competitiveness by improving the quality of products and services as a first-mover advantage~\cite{Porter1995}. At the same time, customer expectations are satisfied because self-regulation organizations (SROs) enforce rules and standards set by themselves to protect consumers, which additionally upholds rights for employees and improve public trust~\cite{Gunningham1997,Short2010}. Self-implemented standards can also span jurisdictions~\cite{Kells1992}. Studies have shown that self-policing across locations makes industry-developed standards more predictable and consistent, and therefore less costly than government regulations~\cite{Gupta1983,Short2010}.

On the other hand, costs of self-regulation activities which are imposed on firms cause them to shift resources away from other activities to achieve compliance. These costs are often justified as a means of improving social welfare, however, also bad factor giving rise to free-rider problem which would cause incredible harm to people, government and finally business~\cite{Hemphill1992}. In detail, in order to be effective, a SRO may set rules for an industry including firms that do not participate in the SRO. These outside firms enjoy the benefits of the regulatory regime without paying any of the costs, as well as those 'bad actors' who also stay outside the system so that they can avoid the rules of the SRO. Such a system is actually unfair to dues-paying businesses, which makes self regulation an inadequate choice for certain industries. This limitation of SROs cannot be ignored. Additionally, self-regulation ineffectively enforces its rules as a punishment tool for governing the private sector, when the problem that massive firms run roughshod over rights seems to be widespread, from India's mining sector~\cite{Indiamining}, to Cambodia's garment industry~\cite{Cambodia}, to the debt buying industry in the United States~\cite{Helleiner2009,Stiglitz2009}. In such cases high proportion of entities are found to be unlawful, the threat or even powerful measures of government regulation can lead to a more effective and stronger enforcement by the SRO, or public-expected results through government direct involvements. This suggests that government oversight or enforcement is indispensable or even final guarantee for public welfare, regardless of the fact by its nature it creates barriers to innovation or competitive entry because of its established norms that only capture current market participants and activities. 

Meaningful regulation schemes are usually driven by a complex mix of internal, external, and reputational motivations~\cite{Benabou2006}. Especially, the nature of intrinsic organizational motivation is central to the definition of the both regulation schemes~\cite{Benabou2006,Short2010}. The above statement of the two regulation schemes initially shows that the amount or proportions of disciplined members or bad actors/wrongdoings in the SROs can be considered as crucial intrinsic motivations to drive meaningful implementations of the two regulations. This is also supported by previous empirical studies involving self-regulation~\cite{Hemphill1992,Demarzo2005,Helleiner2009,Stiglitz2009} that the abundance of disciplined members or bad actors in the SROs is an essence to  curve action modes of the two regulation measures. Note that, not only punitive sanctions but also other tools such as regulatory threats and surveillance can be effective means of regulation enforcement, while measures of the two regulations are amused to be punitive actions in our study for the sake of exploration. This assumption is rational, since punitive enforcement or at least the possibility of it, turns out to be essential to the ultimate success of schemes that incorporate either self-regulation or government regulation~\cite{Short2010}. Therefore, for simplicity, in present paper we can correspondingly define self-regulation and government regulation as cooperator-driven (CP) and defector-driven punishment (DP), within the framework of game theory, by virtue of the important one of their intrinsic motivations: the abundance of bad actors or disciplined members in the SROs. More in detail, implementation of CP and DP are significantly driven by the abundance or proportion of disciplined and undisciplined members in the SROs (we will give concrete form on how the intrinsic motivations quantitatively govern the implementation probability of the punishments in Sec.~\ref{sec:model}) respectively, which is also the definition of the two punishment measures. Meanwhile, it must be stressed that enforcement tools of the two regulations especially self-regulation are diverse in reality. One example of self-regulation is Financial Industry Regulatory Authority (FINRA), which is subject to United States Securities and Exchange Commission (SEC) oversight, and which imposes penalties on bad brokers~\cite{Black2013}. Besides, the other enforcement actions of self-regulation like being excluded from the association and/or making public the accusations are non-negligible~\cite{Hull2004}. In accordance with the proposed definitions of the two different regulations, the regulation issue discussed above can also be well mapped to a pool punishment in which the two different punitive measures ought to be captured.

In reality, SROs operate essentially to protect the interests of individual firms or the industry as a whole, while governments are more concerned for protecting social welfare because they face the pressures from the public all the time~\cite{Castro2011}. That is to say, unlike government regulation, industry-created standards run the risk of advancing commercial interest over public interest. For the SROs, the only way is to move faster (i.~e. make higher quality standards or stricter regulations than do governments) than the government so that they can timely avoid greater benefit losses caused by government regulation or public criticism~\cite{Maxwell2000}. That is why SROs are more willing to nip the irregularities in the bud (or there are few illegal firms to be identified in SROs). However, when high proportion of business actors of one association grow too comfortable accepting and helping to entrench because of a particular kind of lawlessness~\cite{Indiamining,Helleiner2009,Stiglitz2009}, considerable cost of regulations may make the interests of a particular industry and society do not align, the SROs or industry associations will not collaborate to make punishment available, but rather collude to protect vested interests instead of public interests in absence of any external pressure from government stakeholders~\cite{Mayer2010}. Such activities can thus reduce social welfare. Many of these concerns are finally allayed by independent non-profit public-interest organizations such as government oversights and audits which could strongly monitor and enforce rules~\cite{Mayer2010}. This is the essential mechanism for why cooperator-driven punishers (defector-driven punishers) make a decision to exert punishment more when the proportion of cooperators (defectors) is higher in SROs. Relating to our model, this reality suggests that our theoretical hypothesis of the function modes of the two different pool punishment measures are rational and realistic to some extent. It is thus more convinced that government organizations can be theoretically represented by defector-driven punishers while SROs can be perfectly mapped to cooperator-driven punishers.

\begin{figure*}
\includegraphics[width=\textwidth]{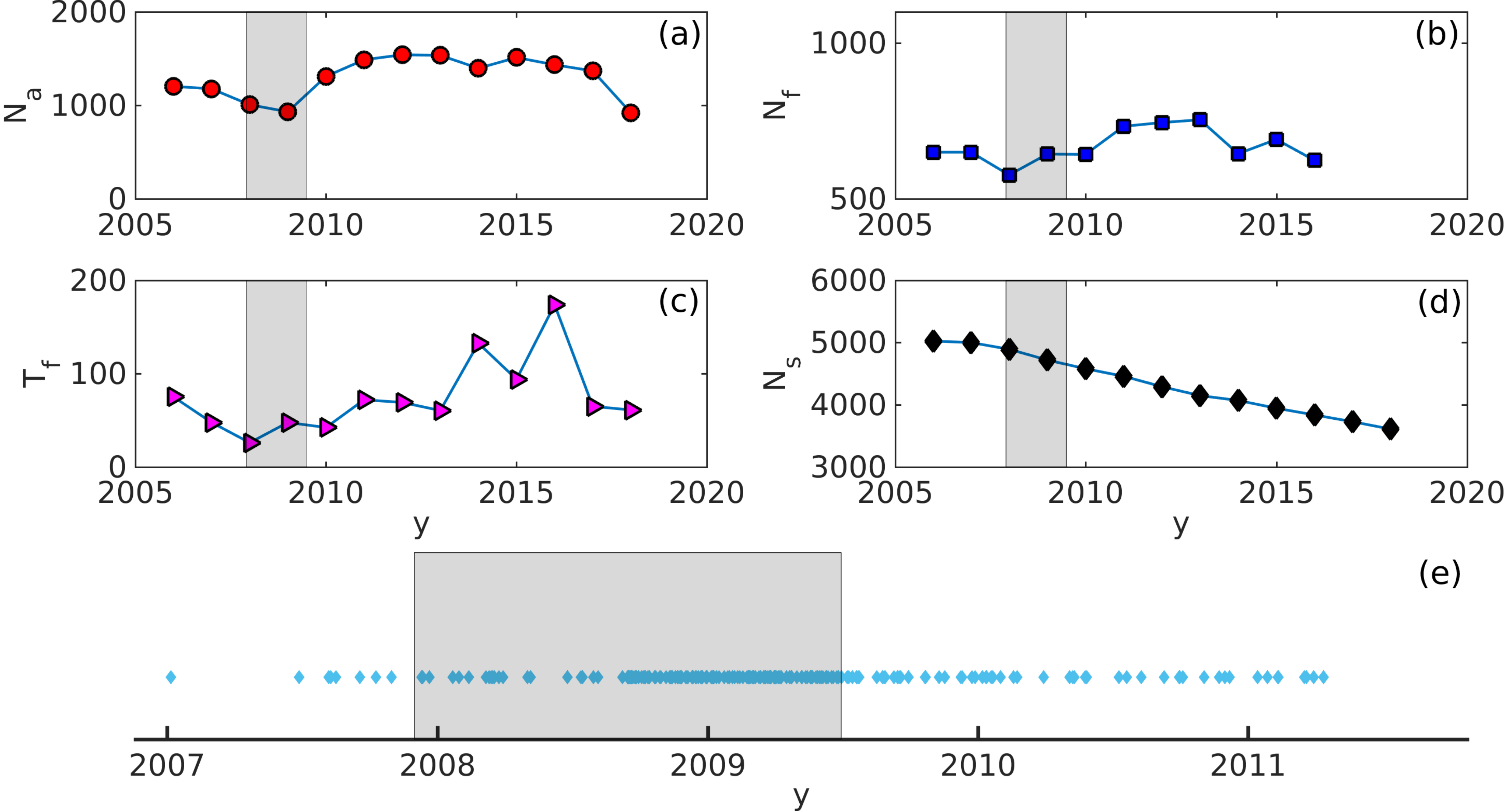}
\caption{Empirical evidences for the regulatory activities by either FINRA~\cite{NASD2006,FINRA2007,FINRA2008,FINRA2009,FINRA2010,FINRA2011,FINRA2012,FINRA2013,FINRA2014,FINRA2015,FINRA2016,FINRA2017,FINRA2018,Statistics} or U.S. Government/The Federal Reserve (The FED)~\cite{Fedtimeline}, which are consistent with the rules made by our model. More in detail, five key statistical characterizing quantities are shown: (a) $N_{a}$ denotes the number of disciplinary actions against firms and their employees, which are brought by FINRA ---the self-regulatory organization (SRO) for brokerage firms doing business with the public in the United States. The disciplinary actions may result in sanctions including censures, fines, suspensions and, in egregious cases, expulsions or bars from the industry~\cite{FINRA2016}. (b) $N_{f}$ denotes the number of fines which represent sanctions exerted by FINRA for rule violations. It should be noted that the number of Finra fines for 2017 and 2018 is still either being counted or even unavailable. (c) $T_{f}$ is the total amount of fines levied by FINRA from individual brokers
and firms. (d) The size of the whole population of registered members (i.e. FINRA-registered firms instead of disciplined members or cooperators) which gradually decreases with time. (e) The regulatory activeness of U.S. Government/The Federal Reserve, in which each diamond represents a press release about regulatory policy or act made by public institutions in U.S. Government/The Federal Reserve. Obviously, the time windows during which denser diamond markers can be observed reveal more frequent involvement of government regulations. In all subfigures, the filled areas represent the duration of 2007-2009 financial crisis (December 2007-June 2009)~\cite{Hulbert2010}. Up to now, we cannot know the accurate frequency or number of cooperators (i.e., disciplined members) or defectors (i.e., undisciplined members) during this financial crisis because not all bad actions cannot be successfully identified and the cost of investigations themselves are huge. However, it must be stressed that the frequency of disciplinary actions against firms which are brought by government or FINRA are positively related to the quantity of undisciplined actions in the course of financial crisis or at other times, respectively. Furthermore, more undisciplined members were 'identified' as forms of bankruptcy, higher non-performing loan ratio or being judicially investigated during the crisis than at other times~\cite{Fedtimeline,Statistics,Volden2012}. It evidently reveals that there exist more identified undisciplined members in the course of financial crisis, which thus correspondingly suggests lower proportion or fewer cooperators at this time because of fewer registered members than before (Figure.~(d)).
}
\label{fig:empiricaldata}
\end{figure*}
An global apocalypse -- the 2007-2009 financial crisis~\cite{Volden2012}, may provides some key hints on how the two punishment measures  intervened along with an increase in bad debts brought about by more bad actors, and what performances they had. Fig.~\ref{fig:empiricaldata} provides further empirical evidences by illustrating five key statistical characterizing quantities (please see the caption of Fig.~\ref{fig:empiricaldata} for more details of these quantities): the number of disciplinary actions against firms and their employees which are brought by FINRA, the number of fines which represent sanctions exerted by FINRA for rule violations, the total amount of fines levied from individual brokers and firms, the number of member firms (i.e. FINRA-registered firms), the regulatory activeness of U.S. Government/The Federal Reserve. At the early stages before 2008, SROs such as National Association of Securities Dealers, Inc. (NASD), New York Stock Exchange (NYSE) or even their combination FINRA can still regulate their members through execution of their enforcement programs such as great sanctions (i.~e. financial penalties)~\cite{FINRA2008,Peirce2015}. Correspondingly, it can be observed in Fig.~\ref{fig:empiricaldata} that self-regulation activities remain flat or even high level outside the filled areas (Fig.~\ref{fig:empiricaldata}(a)-(c)), along with a decrease in government deregulation in finance (see Fig.~\ref{fig:empiricaldata}(e)). However, with persisting push from the loan market, more and more financial companies reached a tacit understanding so as to protect and get vested interests through 'creating' various financial innovations which greatly change the leverage. In such case, many of the companies rabidly opposed any move to make those standards mandatory or to enforce relevant legal standards more vigorously. As a result, self-regulation presented less efficiencies; further leading to approach of the crisis which finally happened along with the bankruptcy of Lehman Brothers and acquisition of Merrill Lynch~\cite{Ivashina2010}. At this moment, the public became rather angry, which made the government begin to police the financial market by means of highly punitive measures against some undisciplined firms, with the assistant of FINRA~\cite{FINRA2008,Volden2012}. Fig.~\ref{fig:empiricaldata}(a)-(c) thus show that a valley in self-regulation interventions from FINRA occurred within the the duration of 2007-2009 financial crisis indicted by the filled areas in which, however denser diamond markers observed in Fig.~\ref{fig:empiricaldata}(e) mean that more frequent involvements of US Government/The FED are released. The regulatory measures by government may be various and not limited to punishments like more stringent acts (like Dodd-Frank Wall Street Reform and Consumer Protection Act) and judicial investigations most of which are actually stimulated by this crisis. Whatever, more frequent involvements of these measures definitely reveal greater willingness of government to regulate the financial firms and their representatives at the heart of the crisis (see Fig.~\ref{fig:empiricaldata}(e)). Moreover, we can observe that considerable government policies are still released afterwards the crisis because policy making is rather time consuming and thus always time-delayed. Taken together, this disastrous process clearly shows a drawing: implementations of self-regulation or cooperator-driven punishment are promoted by abundant disciplined members who are willing to share the risk and costs of execution, while government (i.~e. defector-driven punisher) is more willing to make a strong intervention through defector-driven punishments when the industry is widely eroded by a large amount of bad actors. In any case, the number of FINRA-registered firms gradually decreases with time, which further proves the above conclusion. This classical example further supports the statement that the amount of undisciplined (or disciplined) is the key factor to mainly dominate intervention level of the two different punishments.

It is believed that this important function mode of pool punishment can be found in many other realistic social systems. {\em The key fact that motivated our present work is that, with respect to prosocial pool punishment, the two punishment measures against free-riding may have definitely different performances due to different evolution situations corresponding to different combinations of strategies and different population structures.} For a model of evolution dynamics to capture the real behaviors as accurately as possible, the distinct probability to curve  establishment/enforcement degree of the two different punishment measures must be taken into account in framework of pool punishment.

In addition, a widely-applied game framework -- public good game (PGG) provides a good theoretical framework to concern the public welfare in presence of pool punishment. Group-like structure in PGG is in favor of function of pool punishment, even though the overlap between different game groups depends on the network structures. Another reason for PGG model being the first choice as the present game framework is that its rules are very closed to the operate modes of many modern money-seeking organizations such as banks, profit funds or listed companies: attracting capital and then sharing investment gains together.

To our knowledge, in the current literature, there is no work on evolutionary dynamics which takes into account issues of self-regulation and government regulation. In general, both types of regulations exist, and the question is how important they are in governing the evolutionary dynamics in different situations. In this paper we propose a general evolutionary game model to capture the two distinct punishment measures: defector-driven and cooperator-driven punishment, in the framework of PGG. Next, the evolution dynamics in six different evolutionary situations on well-mixed population and networked population is detailedly treated by mean-field theory and extensive agent-based simulations respectively. Moreover, our developed semi-analytical method is able to yield accurate estimations of the boundaries between the phases of full defection and nondefection. We provide several physical understandings of the basic evolutionary dynamics for different situations on the two populations, through a detailed statistical analysis, uncovering the favorable conditions under which cooperation prevalence with abundant punishers can arise. 

Three striking phenomena are uncovered in this study. One is that networked structure is overall more favorable for punishers and further for cooperation because of network reciprocity and mutualism between punishers and cooperators who do not punish defectors. The second phenomenon  is that, for networked populations, cooperator-driven punishment is a more efficient measure to confer nondefectors evolutionary advantages; however, on infinite well-mixed populations the mix of the two punishments is a better choice to achieve a desirable evolutionary outcome. 
Finally, of particular interest is that, for networked population, an optimal intermediate range of feedback sensitivity for prevalence of punishers is identified. We give a clear physical picture to help us understand how this phenomenon happens, through a detailed statistical analysis of spatial pattern formations in different evolution situations.  

The paper is organized as follows. We firstly give a detailed description of our model in Sec.~\ref{sec:model}. In Sec.~\ref{sec:results}, in turn, we firstly implement our model for six different evolutionary situations on infinite well-mixed population through theoretical approach, and then extend our study to networked populations by means of agent-based simulations. Finally, we present discussions and conclusions with an outlook in Sec.~\ref{sec:discuss}.

\section{Model}
\label{sec:model}

In our model, regulations or policing involvements are directly assumed to be punitive measures so that one can design feasible model for analytical and numerical studies of the effectiveness of two different punishments in different evolutionary situations. In reality, regulations may create costs as well as benefits from the increasing levels of disciplined behaviors, so one should consider the cost of punishment in the model in addition to the punishment fine to quantify the punitive effects in terms of high-order benefits (i.e., the emergence and persistence of cooperators or prosocial punishers). In the matter of regulation of industry, the key players in the promotion of public interests have always been businesses, SRO, government and consumer advocates; which should be considered as basic ingredients or strategies in our present model. Accordingly, there are four strategies within the framework of PGG: traditional cooperation (i.~e. nonpunishing cooperation) (TC), defection (D), cooperator-driven punishment (CP) and defector-driven punishment (DP); and correspondingly four types of individuals: traditional cooperators (nonpunishing cooperators), defectors, cooperator-driven and defector-driven punishers. Of particular note here is that the execution probabilities of cooperator-driven punishment (CP) and defector-driven punishment (DP) are mainly determined by the fractions of cooperators and defectors within the game group, respectively; which is also the concrete definition of the two punishment measures in our model. As another punishment measure, traditional punishment (TP) is usually implemented when there is at least one defector in the group, and thus widely-adopted. However, we have checked that this punishment strategy remains rather vulnerable and negligible, especially when CP or DP is present. It reveals that CP and DP are effective ways to regulate defectors in the institution rather than TP, at least within the framework of the present model. Hence this punishment is not considered in our present model. For simplicity, in our model the punishment mechanism is only stated on prosocial punishment, i.e., punishers adopt cooperation strategy before punishing defectors, which means that no other mechanisms such as antisocial punishment~\cite{Rand2010,horne2016,szolnoki2017} or selfish punishment~\cite{cui2014} are captured. 

According to our game rule, in the population each player $i$ selects $G_{i}-1$ individuals from the population to form a game group in which each group member can simultaneously play PGG with other group members, by holding the same strategy. In detail, in the game group each cooperator makes a contribution of $1$ to the public good, while defectors contribute nothing. Subsequently, the sum of all the contributions in the group is multiplied by the synergy factor $1<r<G_{i}$, which quantifies synergistic effects of cooperation. Then the resulting amount is equally shared among all members in the group. After the intervention of punishments, there are two different cases to be considered for a member $i$ adopting cooperation strategy: (1) the payoff of $i$ will be $\Pi_{P}^{g}=r(N_{TC}+N_{CP}+N_{DP})/G_{i}-1-N_{D}\alpha /n_{P}$ if she/he carries out punishment upon defectors in the group at a probability $g$, as either a cooperator-driven punisher (CP) or a defector-driven punisher (DP); (2) otherwise the payoff of $i$ is $\Pi_{P}^{g}=\Pi_{TC}^{g}=r(N_{TC}+N_{CP}+N_{DP})/G_{i}-1$ which is also the payoffs of traditional cooperators in the group. Herein $N_{TC}$, $N_{D}$, $N_{P}$, $N_{CP}$ and $N_{DP}$ are, respectively, the number of traditional cooperators, defectors, punishers, cooperator-driven punishers, and defector-driven punishers in the group. Therefore, $N_{P}=N_{CP}+N_{DP}$. Meanwhile it must be stressed that $n_{p}$ is the number of punishments exerted by punishers rather than the total number of the two types of punishers, while $\alpha$ is the punishment fine that each defector in the group incurs in presence of punishment. In the case that $i$ is a defector, $\Pi_{D}^{g}=r(N_{TC}+N_{CP}+N_{DP})/G_{i}$ if $n_{P}=0$, or else $\Pi_{D}^{g}=r(N_{TC}+N_{CP}+N_{DP})/G_{i}-\alpha$. Importantly, the values of $\alpha$ are kept the same for cooperator-driven and defector-driven punishment so as to not give either a default evolutionary advantage or disadvantage.

More precisely, according to the definition of DP and CP, the probability (i.e., $g$) that the two types of punishments are implemented by corresponding punishers is dominated by the fractions of different strategies in the group, in the following specific way: 
\begin{equation}
\begin{cases}
g_{DP}=A\frac{N_{D}}{G}, \\
g_{CP}=A\frac{N_{C}}{G}.
\end{cases}
\label{eq:asfunztion1}
\end{equation}
where $g_{CP}$ and $g_{DP}$ indicate the probability at which cooperator-driven and defector-driven punisher to carry out an exertion, respectively. $N_{C}=N_{TC}+N_{CP}+N_{DP}$ denotes the total number of nondefectors (including traditional cooperators, cooperator-driven and defector-driven cooperators) in the group. Parameter $A \in [0,~1]$ quantifies punishers' feedback sensitivity. In more detail, larger value indicates more sensitive punishers and larger difference between the two types of punishers in terms of their behavior modes, and thus more exerted punishments with respect to the same fractions of defectors (nondefectors) in the group. 

Furthermore, Eq.~\ref{eq:asfunztion1} reveals that the two types of punishers have two opposite feedback modes. More precisely, defector-driven punishers prefer to implement punishment to bring the population back from brink of collapse caused by abundant defectors, regardless of vast amount of punishment costs which could greatly reduce their payoffs. In contrast, cooperator-driven punishers are more likely to take actions insofar as a good number of nondefectors who can share the costs induced by the punishment, in purpose of reserving their payoffs firstly. To some extent DP and CP construct a new kind of social dilemma with traditional cooperators other than the traditional dilemma consisting of traditional cooperators and defectors. As 'prudent' guys, traditional cooperators play the role of second-order free riders because they consequently preserve higher payoffs than those punishers, with doing nothing to fight against wrong-doers.

In what follows, we will explore the evolutionary dynamics through both mean-field theory concerning infinite well-mixed situation and agent-based simulations in structured populations under various parameter conditions. We have checked that localized interactions on the square lattice with $\langle k\rangle=4$ can lead to obvious inconsistencies between analytical predictions and simulations, by means of  both great evolutionary advantages conferred to the nondefectors and considerable critical slowing down of the system. Moreover, eliminating  critical slowing down of the system with fewer connections is difficult and rather time-consuming. Therefore the structured population in our study is instead curved with a regular lattice with mean degree $\langle k\rangle=6$ and with periodic conditions, with reserving novel findings from our model. On the network of size $N$, an over-lapping game group contains all the nearest neighbors of the focal individual in addition to itself, where each individual simultaneously plays the game. At the same time, each player $i$ holds a PGG played by $G_{i}=k_{i}+1$ group members (together with all his/her neighbors), in addition to participating in $k_{i}$ games initiated by his/her neighbors; where $k_{i}$ is the number of focal individual's neighbors (i. e., the degree). Therefore, each individual $i$ simultaneously plays $k_{i}+1$ PGGs by holding the same strategy.

Furthermore, in the structure population Monte Carlo simulation (MC) is employed to update the strategies of players. And random sequential updating is implemented as the updating scheme to control the evolution. Initially each player fixed on the networks is randomly and independently designated as either a traditional cooperator, a defector, a cooperator-driven or defector-driven punisher. Each time step consists of $N$ following steps such that every player can update its strategy once on average: (1) A randomly selected player $i$ accumulates its overall payoff $\Pi_{s_i}$ by playing the PGG in all the $G_{i}$ groups as a member, where $\Pi_{s_i}$ is thus the sum of all the payoffs
$\Pi_{s_i}^{g}$ acquired from each individual group. The randomly-chosen nearest neighbors $j$ also obtains its overall payoff $\Pi_{j}$ in the same way. (2) Then $i$ simulates the strategy of $j$ with probability given by fermi study function $W_{j\longleftarrow i}=1/[1+\exp((\Pi_{i}-\Pi_{j})/\kappa)]$. The function implies that players owning higher payoffs are advantaged, while adoption of strategy of a player performing worse is still possible. $\kappa$ curves the noise of the uncertainty in the adoption. Without loss of generality we set $\kappa=0.1$ throughout this paper. The simulations are performed until the system reaches a stationary state, i.~e., the populations of different strategies become time independent or defectors go extinct.

The final densities of different strategies are averaged over $200$ independent realizations to insure a low variability. The size of the network is $N=200\times 200$.

\section{Results}
\label{sec:results}
Before presenting the main results, we should state that the focus of the present study is on evolutionary dynamics under various parameter conditions for six different combinations of strategies: D+CP, D+DP, TC+D+CP, TC+D+DP, D+CP+DP, and TC+D+CP+DP. Both analytical treatment based on mean-field theory and agent-based simulations for networked population are employed to enable a full exploration. Correspondingly, for an available comparison, there are two parts to be presented in this section: the first is for analytical predictions from infinite well-mixed populations while the second is for networked population embedded on a regular lattice network.

\subsection{Infinite well-mixed populations}
\label{subsec:infinite}

The calculations, reported in Appendix A, give the following results for different evolutionary situations, which theoretically provide a complete picture of the model behavior.

\begin{figure*}
\includegraphics[width=\textwidth]{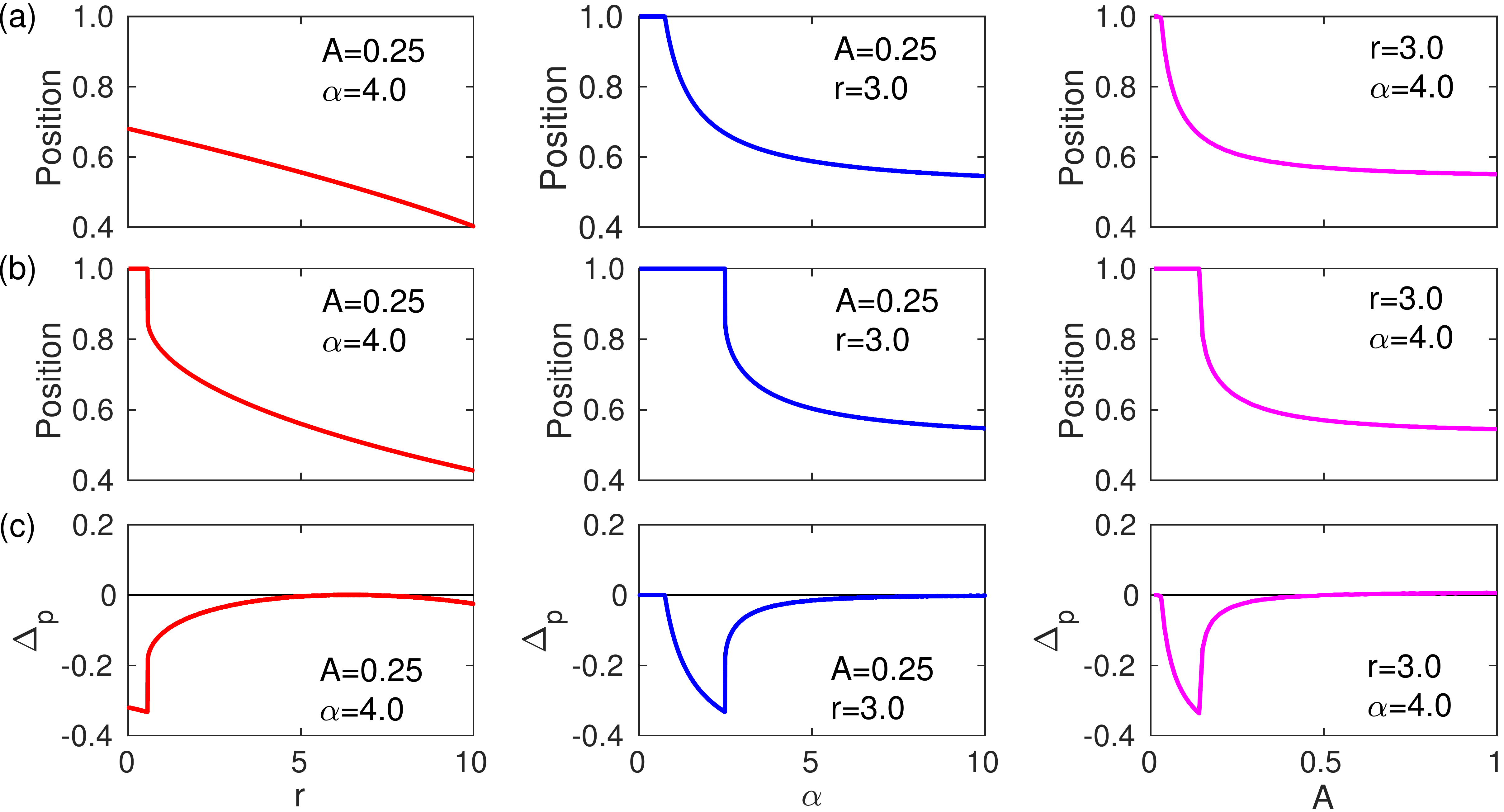}
\vspace{0cm}
\caption{Illustrations of the position of coexistence state of two strategies under various parameter conditions. The position value $p_{CP}$ ($p_{DP}$) indicates the position of intermediate unstable state (i.e., the open circles illustrated in Figs.~\ref{fig:gtwodcp} and \ref{fig:gtwoddp} of Appendix.~B) in the case D+CP (D+DP), which can be used to estimate the advantages of punishment strategy. More specifically, smaller $p_{CP}$ or $p_{DP}$ is, more likely the system would be to reach the sate of full punishment (FP). (a) Top panels is for the case D+CP; while (b) is for the case of D+DP. Specific values of other parameters are depicted in each subplot. (c) illustrates the difference  $\Delta_p=p_{CP}-p_{DP}$ between two position values so that we can judge which punishment is more effective in enlarging the attractive range of FP, i.e., promoting the public cooperation.
}
\label{fig:position}
\end{figure*}

Accordingly, Fig.~\ref{fig:position} gives a complete picture about how the effects of synergistic effects of cooperation, punishment fine and individual sensitivity on evolution direction of the system. Overall, larger $r$, $\alpha$ and $A$ can confer the punishers more evolutionary advantages, leading to the decrease of position values $p_{CP}$ or $p_{DP}$. In particular, this result reveals that more sensitive punishers are able to protect the population from being corroded by defectors through more available punishment. At the same time, second and third columns of panels (a) and (b) indicate that the promoting effects from the punishments and individuals' sensitivity have an upper limit, because attractive range for full punishment (FP, i.~e. the system is full of punishers) identified by the position of the intermediate state remains almost invariable. Negative valley in terms of $\Delta_p$ becomes visible when the parameter condition is not so desirable for the punishers (i.~e., $r$, $\alpha$ and $A$ are small), suggesting that CP is more effective than DP in suppressing defectors under well-mixed conditions. As the three parameters getting larger, there is no significant difference between the two types of punishment with respect to promoting public goods; and $\Delta_p$ is negative but very close to zero as a result. Whatever, none of the punishers can occupy a superior position in face of defectors, which is confirmed by the phenomena $p_{CP}>0.5 (p_{DP}>0.5)$ in Figs.~\ref{fig:position}(a) and (b).

\begin{figure*}
\includegraphics[width=\linewidth]{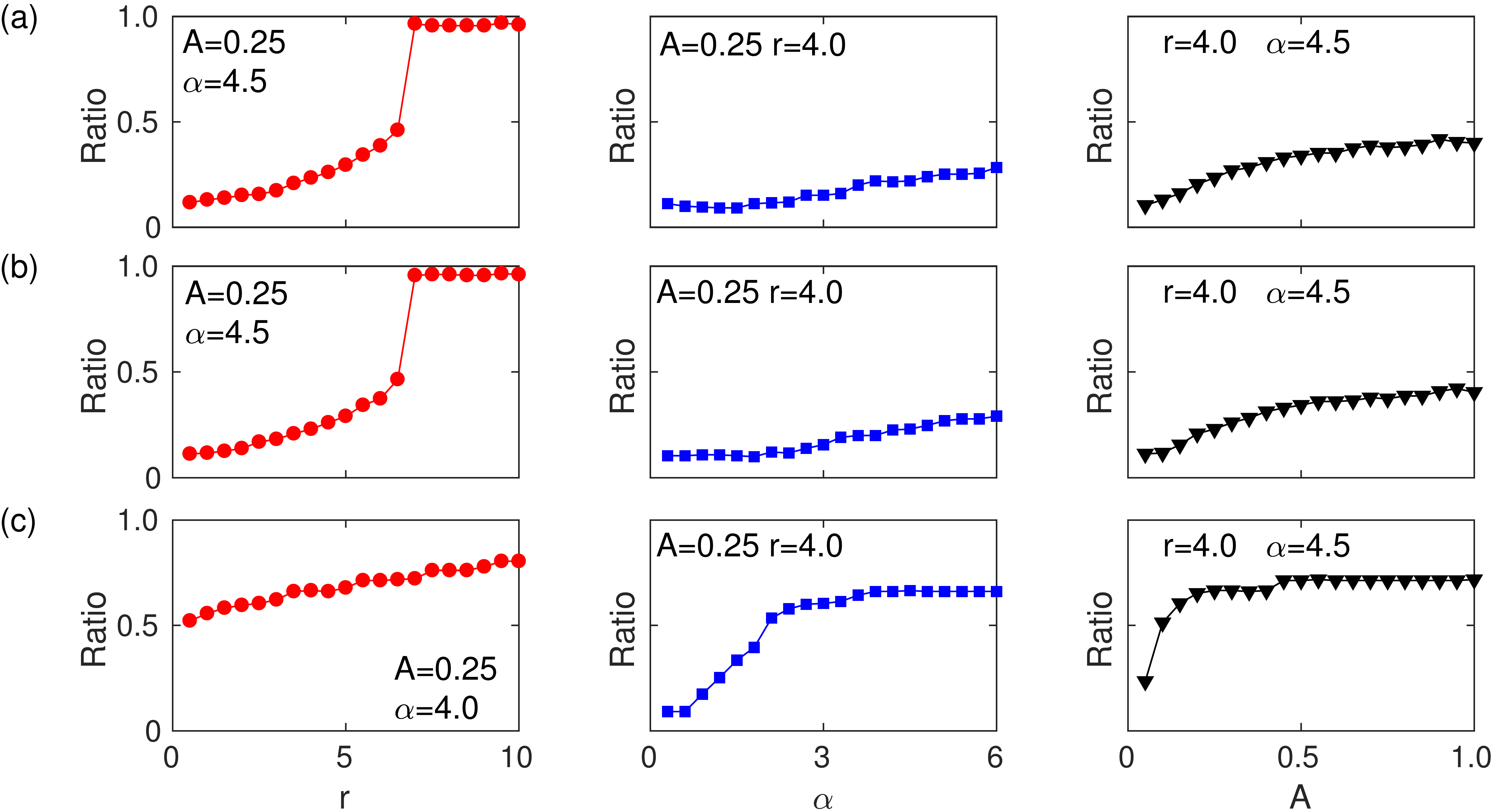}
\vspace{0cm}
\caption{The ratios of the areas of attraction basins of SP state versus $r$, $\alpha$ or $A$; for three different evolution situations: (a) TC+D+CP, (b) TC+D+DP and (c) D+CP+DP.
}
\label{fig:rationofour}
\end{figure*}

Next, we shift our attention to the three-strategy situations: TC+D+CP, TC+D+DP, and TC+CP+DP. Fig.~\ref{fig:rationofour} reveals the monotonous effects of facilitating the advantages of nondefectors of synergy effect, punishment fine and feedback sensitivity, through presenting the ratios of attraction basins of SP under various parameter conditions (please see Figs.~\ref{fig:threecddp}, \ref{fig:threecdcp}, \ref{fig:threeccpdp} and corresponding descriptions for more detailed information of attraction basin patterns for the three-strategy situations). While an obvious saturation effect can be found in the case D+CP+DP, where the ratios reach a limited intermediate value with increasing punishment fine or feedback sensitivity (see the second and third subplot in Fig.~\ref{fig:rationofour} (c)). Secondly, comparatively higher ratios suggest that mix of CP and DP works better in sustaining the public goods without involvement of second-order free riders i.e., traditional cooperators; which is in accordance with the conclusion from the empirical studies that combination of self-regulation and government oversight lead to a better performance in improving market track~\cite{Demarzo2005,Sharma2010}. On the other hand, it can be noticed in the cases with sole punishment that the performance of synergy effect goes through a sharp transition from being unimpressive to being rather remarkable near the point $r=G$ above which nondefectors (ND) dominate while all defectors die out, which is more or less in agreement with the findings of the previous works~\cite{Rand2010}. 

\begin{figure*}
\includegraphics[width=\linewidth]{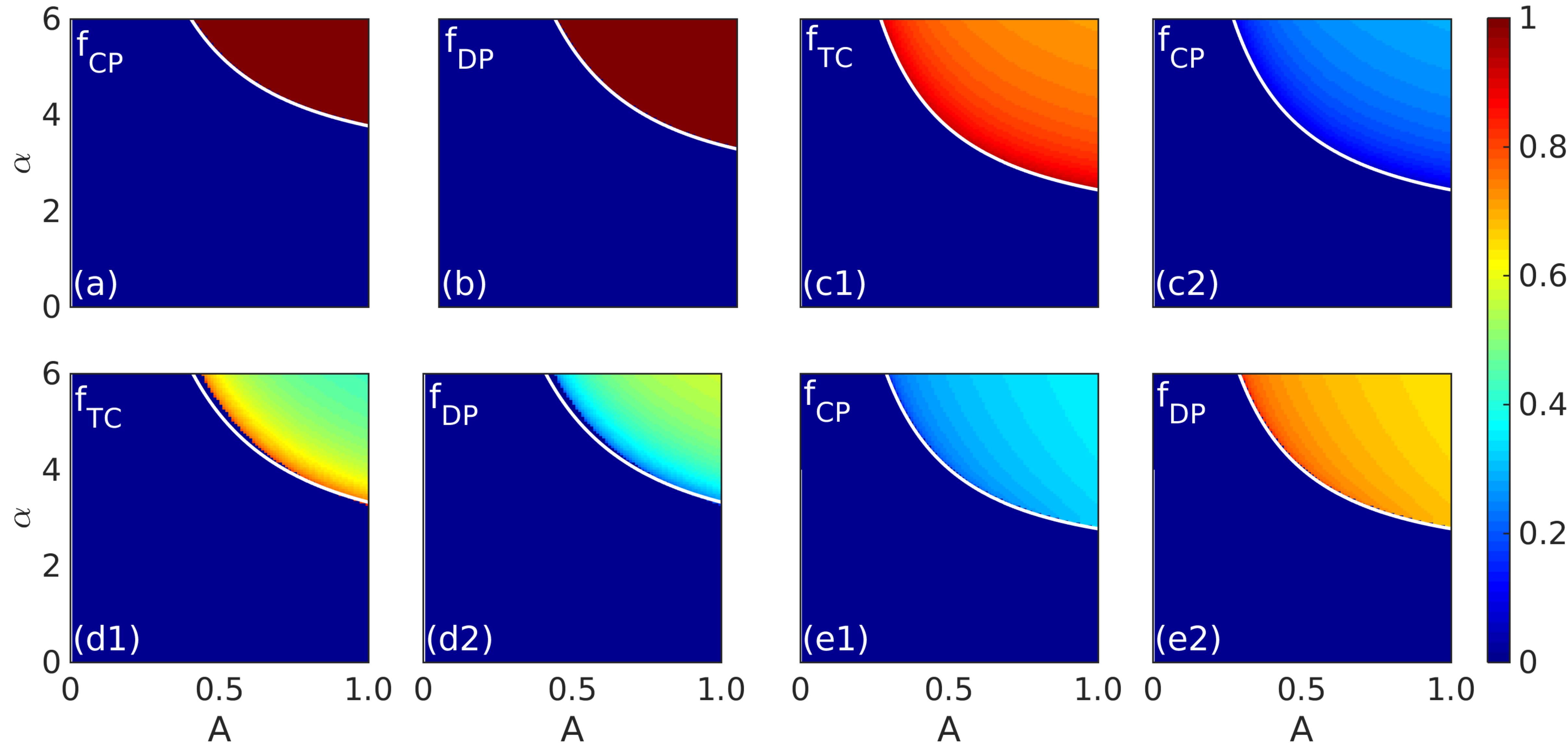}
\vspace{0cm}
\caption{The analytical dependence of the final steady fractions of different nondefective strategies on both $A$ and $\alpha$ for different evolutionary situations: (a) D+CP, (b) D+DP, (c1) and (c2) TC+D+CP, (d1) and (d2) TC+D+DP, (e1) and (e2) D+CP+DP. The initial conditions for each case are: (a) $f_{CP}(0)=0.54$, (b) $f_{DP}(0)=0.54$, (c1) and (c2) $f_{CP}(0)=0.45$ and $f_{TC}(0)=0.22$, (d1) and (d2) $f_{DP}(0)=0.53$ and $f_{TC}(0)=0.05$, (e1) and (e2) $f_{CP}(0)=0.3$ and $f_{DP}(0)=0.25$. Correspondingly, the values of $f_{s}$ used to semi-analytically estimate the boundary lines are (see Appendix~C for further details): (a) $f_{CP}=0.54$, (b) $f_{DP}=0.54$, (c1) and (c2) $f_{TC}=0.22$ and $f_{CP}=0.356$, (d1) and (d2) $f_{TC}=0.05$ and $f_{DP}=0.492$, (e1) and (e2) $f_{CP}=0.3$ and $f_{DP}=0.25$. In all cases, the value of synergy factor is $r=4.0$. 
}
\label{fig:analysisnofour}
\end{figure*}

\begin{figure*}
\includegraphics[width=\linewidth]{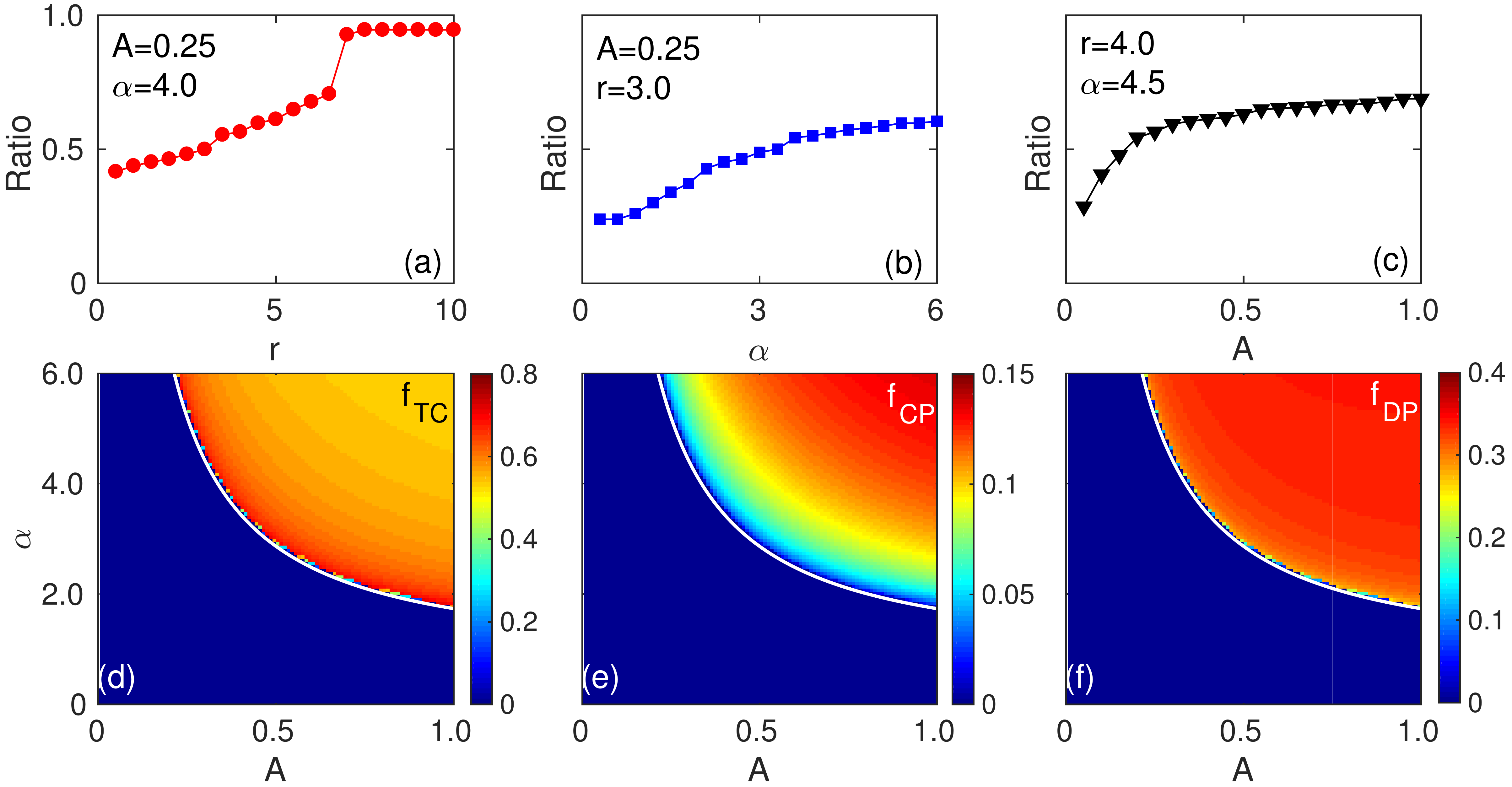}
\vspace{0cm}
\caption{Comprehensive understanding of the evolutionary dynamics for the evolutionary situation TC+D+CP+DP. Top panels: The volume ratios of attraction basins of SP state versus $r$, $\alpha$ or $A$, where the values of other parameters are given in the subplots. Bottom panels: The analytical dependence of the final steady fractions of three nondefection strategies on both $A$ and $\alpha$, along with the semi-analytical solid boundary lines. The initial condition is $\rho_{S}(0)=0.25$ (S$\in$\{TC, D, CP DP\}). Correspondingly, the values of $f_{s}$ used to semi-analytically estimate the boundary line are: $f_{TC}=f_{DP}=f_{CP}=0.25$. The other parameter is $r=4.0$.
}
\label{fig:ratioanalysisfour}
\end{figure*}

Furthermore, Figs.~\ref{fig:analysisnofour} and \ref{fig:ratioanalysisfour} together provide a comprehensive picture of strategy fractions in the parameter plane of ($A$, $\alpha$) for six different evolution situations, as well as semi-analytically estimated boundary lines and volume ratios of attraction basins of segment punishment (SP, the state where punishment is found to be coexisted with other strategies) for the case TC+D+CP+DP. Especially, $f_{TC}$, $f_{CP}$ and $f_{DP}$ represent the final steady fractions of traditional cooperators, cooperator-driven punishers and defector-driven punishers in the population, respectively. Most obviously, the regions of ND phase are found to be in the top-right corner, which can be considered as a support for the observed monotonous effects of facilitating the advantages of nondefectors of $r$, $\alpha$ and $A$ from another perspective. Since there are much more cost of punishment defectors have to bear, induced by both more frequent exerted punishments and larger cost of one punishment. It can also be expected that the semi-analytical approach relying on well-mixed assumption can give 'perfect' boundaries to distinguish SP phases from FP phases. Cooperator-driven punishers are in a disadvantaged position in face of traditional cooperators or even defector-driven punishers (more traditional cooperators or defector-driven punishers exist in the final state, see Figs.~\ref{fig:analysisnofour} (c1)~(c2), (e1)~(e2), and Figs.~\ref{fig:ratioanalysisfour}~(d)-(f)). Further support for requiring a non-trivial interplay between defector-driven and cooperator-driven punishers in promoting the public cooperation is also verified in Fig.~\ref{fig:analysisnofour} (e1) and (e2) where a larger region of ND phase is existed. At this point, CP is frequently enforced when nondefectors dominate the population while DP efficiently works on condition that defectors become the majority. Therefore sufficient available punishments are always provided in spite of the strategy abundances. Especially through a deep comparison of the illustrations in Fig.~\ref{fig:rationofour} with the ratios in Fig.~\ref{fig:ratioanalysisfour}, we can say the transition in terms of ratio is mainly due to synergistic effects from cooperation, instead of punishments themselves. In other words, punishment fine and feedback sensitivity are the parameters that mainly govern the performance of punishment measures in an infinite well-mixed population. Finally, combining with the two figures, we can find traditional cooperators are traditionally held responsible for preventing the dominance of cooperator-driven and defector-driven punishers, in accordance with conclusions from previous studies~\cite{panchanathan2004,Sigmund2010,perc2012}.

\subsection{Networked populations}
\label{subsec:networked}

The study of present model under infinite well-mixed condition reveals the monotonous effects of synergistic effects, punishment fine and feedback sensitivity in facilitating the public cooperation. In accordance with the conclusions from few previous empirical studies involving internet co-regulation~\cite{Marsden2011}, combination action of CP and DP gives better performance. Besides, a non-negligible deviation from the reality is that CP is always outperformed by DP or TC. However, in reality, interactions among individuals are not typically random but rather highly structured, i.~e., each individual has a fixed neighborhood to some extent~\cite{szabo2007,Perc2013,Newman2018}. Taking this realistic factor into consideration, we find some novel and counterintuitive results which has not been uncovered in a well-mixed population.

\begin{figure*}
\includegraphics[width=\linewidth]{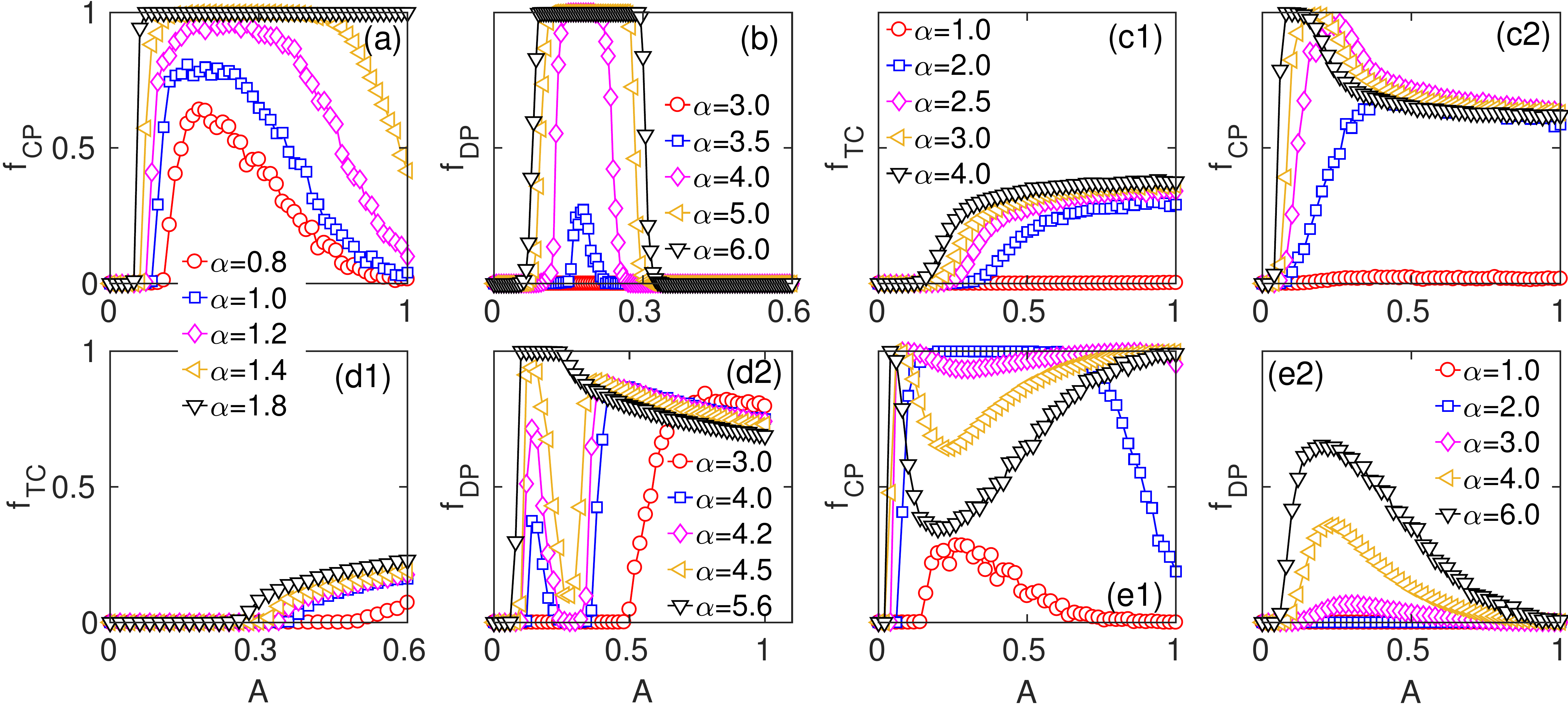}
\vspace{0cm}
\caption{Final steady fractions of different nondefective strategies as function of $A$ for networked populations, where the results for five different evolution situations are respectively illustrated: (a) D+CP, (b) D+DP, (c1) and (c2) TC+D+CP, (d1) and (d2) TC+D+DP, (e1) and (e2) D+CP+DP. In all cases, $r=4.0$.
}
\label{fig:functionanofour}
\end{figure*}

\begin{figure*}
\includegraphics[width=\linewidth]{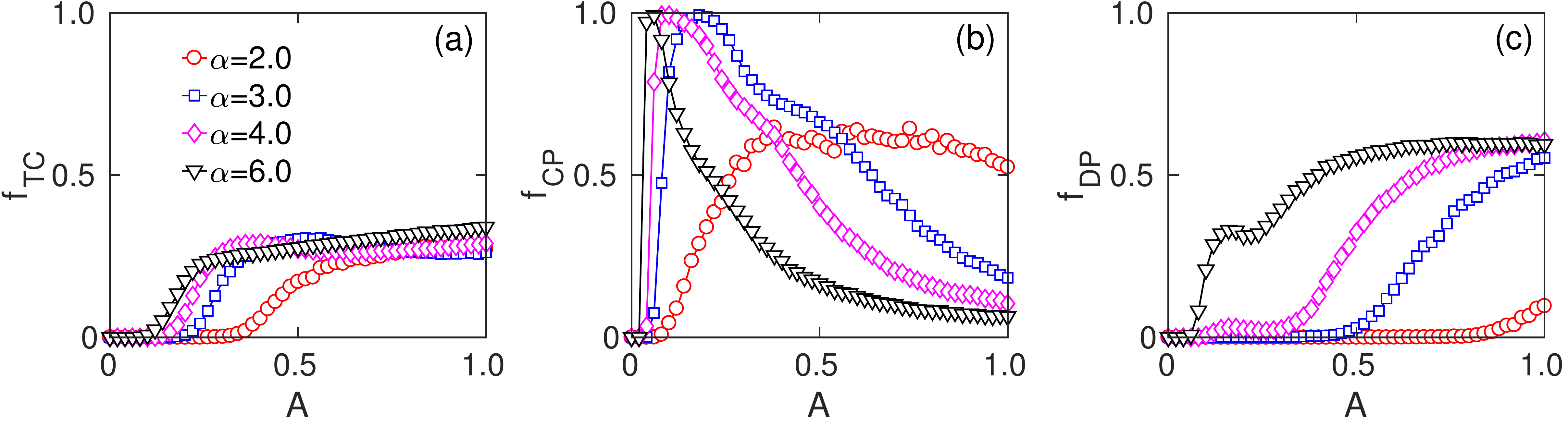}
\vspace{0cm}
\caption{Final steady fractions of different nondefective strategies as function of $A$ for networked populations in presence of four strategies: TC, D, CP and DP. The value of synergy factor is $r=4.0$.
}
\label{fig:functionafour}
\end{figure*}

Firstly, Figs.~\ref{fig:functionanofour} and \ref{fig:functionafour} together exhibit what role feedback sensitivity plays in governing the evolution dynamics in six different evolution situations. Of particular interest is that an optimal intermediate range of the sensitivity $A$ for prevalence or even complete dominance of punishers can be found under suitable parameter conditions in each situation. It is definitely different from what happens in well-mixed populations. More specifically, in the case that cooperator-driven punishers face defectors alone, the peaks of $f_{CP}$ become wider until a threshold of punishment fine above which it exhibits an monotonic increase (see Fig.~\ref{fig:functionanofour}(a)). After the introduction of traditional cooperators, more sharp peaks of $f_{CP}$ can be observed for large $\alpha$, along with a shift of these peaks toward smaller $A$ with increasing punishment fine (see Fig.~\ref{fig:functionanofour}(c2)). In contrast, the emergence of optimal intermediate ranges of $A$ for dominance of defector-driven punishers is instead facilitated by large punishment fine for which either two peaks of $f_{DP}$ are present when traditional cooperators are additionally introduced (see Fig.~\ref{fig:functionanofour}(b) and (d2)). In the cases considering sole punishment, the positions of peaks remain more or less independent of whether TC's intervention. However, the optimal intermediate ranges of feedback sensitivity $A$ would shrink (see Figs.~\ref{fig:functionanofour}), which still holds in the cases with mix of punishments (see Figs.~\ref{fig:functionanofour}(e1)~(e2) and Figs.~\ref{fig:functionafour}(b)~(c)). This arises from the fact that punishers would suffer from second-order free-riding behaviors, which also contributes to monotonic increase of traditional cooperators with the growth of punishers (Figs.~\ref{fig:functionanofour}(c1) and (d1), and Fig.~\ref{fig:functionafour}(a)). Additionally, a key hint for competitive relationship between the two types of punishers is also revealed by the results presented in Fig.~\ref{fig:functionanofour}(e1) and (e2) that positions of population peaks of DP correspond rightly to the valleys of CP population in case that punishment fine is large enough. Figs.~\ref{fig:functionanofour}(e1)~(e2) and Figs~\ref{fig:functionafour}(b) also show that stronger punishment indicated by higher $\alpha$ does not always mean higher achievable levels of CP in the population as both punishment measures are taken together. There is also an optimal intermediate ranges of $\alpha$ for the prevalence of CP individuals. In presence of all four strategies, peaks of $f_{CP}$ rather than $f_{DP}$ still occur for large $\alpha$ (Figs~\ref{fig:functionafour}(b) and (c)). Finally, by comparing Fig.~\ref{fig:functionanofour}(a) with (b) or Figs.~\ref{fig:functionanofour}(c1)(c2) and (d1)(d2), we note that CP is more efficient than DP with respect to maintaining public goods, by promoting a wider range of $A$ for prevalence of punishers with smaller punishment fine.

\begin{figure*}
\includegraphics[width=\linewidth]{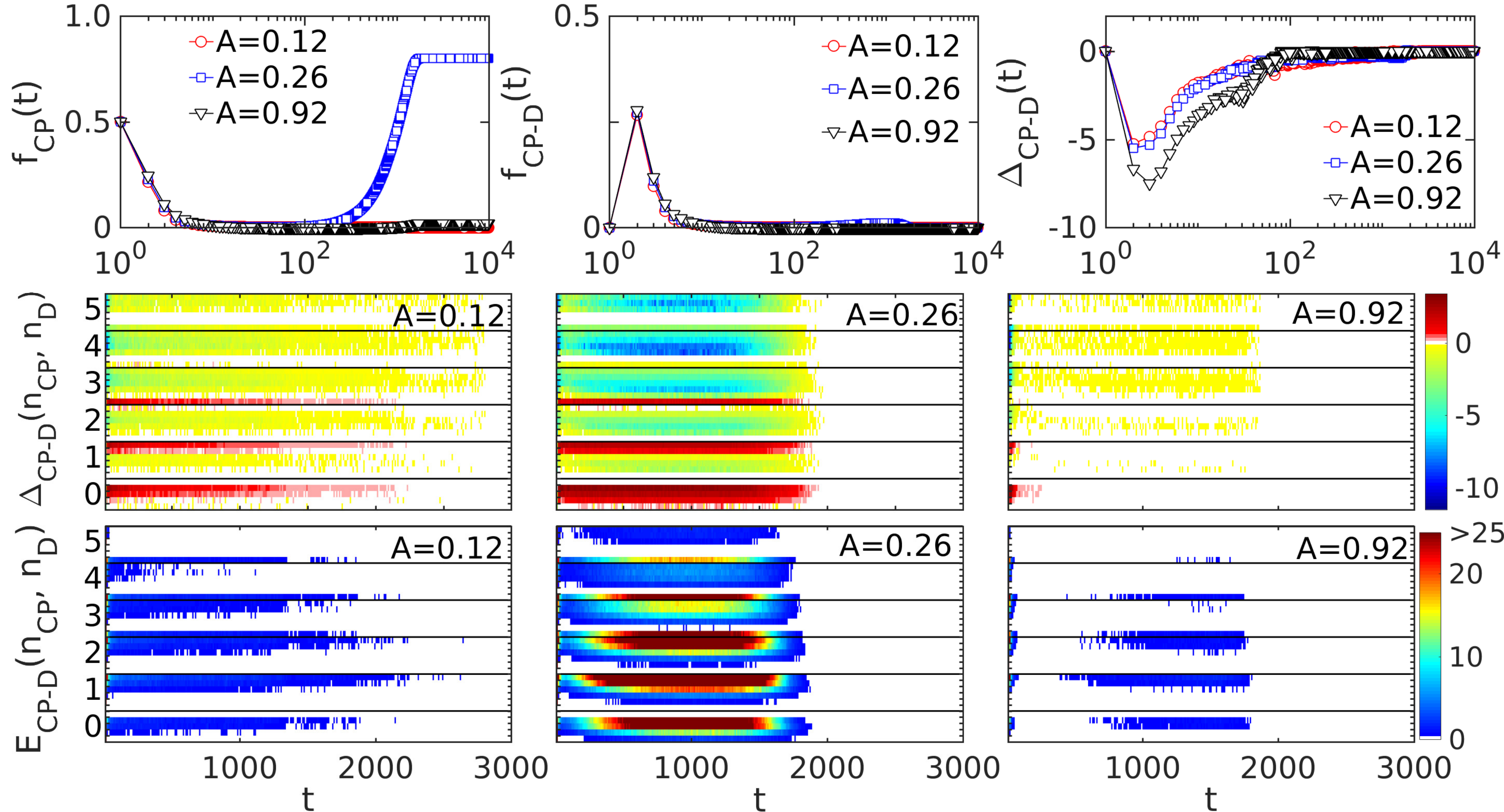}
\vspace{0cm}
\caption{Understanding evolutionary dynamics of the networked populations for three representative values of $A$: (a) $A=0.14$, (b) $A=0.26$ and (c) $A=0.92$, in case D+CP. Five key statistical characterizing quantities are shown for $r=4.0$ and $\alpha=1.0$: the fractions of cooperator-driven punishers $f_{CP}(t)$ (the first subplot in the top panels); the fractions of edges connecting the punishers to defectors $f_{CP-D}(t)$ (the second subplot in the top panels) which are normalized by the total number of edges in the networks; the mean payoff gap between a cooperator-driven punisher and its connected defector among whom an imitation process happens $\Delta_{CP-D}(t)=\sum\limits^{E^{'}_{CP-D}(t)}(\Pi_{CP}-\Pi_{D})/E^{'}_{CP-D}(t)$ (the third subplot in the top panels, and $E^{'}_{CP-D}(t)$ represents the number of total occurrences of imitation process among CPs and Ds at time t); the payoff-gap spectrum ($\Delta_{CP-D}(n_{CP}, n_{D})$) of $\Delta_{CP-D}(t)$ for $36$ different neighborhood states which is determined by the number of cooperator-driven punishers can be found in their neighbors (the middle panels), where $n_{CP}$ ($n_{D}$) represents the number of cooperator-driven punishers can be found in neighbors of a cooperator-driven punisher (defector); the number spectrum of $36$ different states of edges $E_{CP-D}(n_{CP}, n_{D})$ whose states are determined by the number of cooperator-driven punishers the two ends respectively have. In detail, the six numbers marked on the y label indicate the possible number of neighbors who are found to be cooperator-driven punishers for a cooperator-driven punsiher. There are thus six scales for each number (grid box), each of which respectively indicates the numbers of cooperator-driven punishers the defective neighbor has (from bottom to top: the number of punitive neighbors of this defector is 0-5), rising to $36$ scales in total.
}
\label{fig:rounddcp}
\end{figure*}

The microscopic mechanism behind the reported optimal intermediate feedback sensitivity in the case D+CP is rather uncovered by the behaviors of different key statistical characterizing quantities presented in Fig.~\ref{fig:rounddcp}. In particular, the spectrum of payoff gaps $\Delta_{CP-D}(n_{CP},~n_{D})$ and different states of edges $E_{CP-D}(n_{CP}, n_{D})$ presented in Fig.~\ref{fig:rounddcp} reveal more detailed information about the spatial pattern formations. When $A$ is small, unresponsive CP individuals would exert too few punishment on defectors within the groups, leaving the expand opportunities to defectors. The fractions of cooperator-driven punishers thus decrease to zero. Conversely, for large $A$ cooperator-driven punishers are too sensitive to punish too many defectors of different groups, which could greatly reduces punishers' payoffs (see the third plot in the top panels of Fig.~\ref{fig:rounddcp}) especially those with fewer than three connected punishers (Fig.~\ref{fig:rounddcp}), and further rise to shrink of their formed islands. This implies that optimal feedback sensitivity to maximize CP population should be intermediate. Under such parameter condition, cooperator-driven punishers could not only defeat defectors through sufficient and strong punishment (large $\alpha$) but also maintain competitive advantages (especially those owning less than three neighbors of the same strategy, see the middle panels in Fig.~\ref{fig:rounddcp}) at the borders of CP clusters so as to finally expand permanently by absorbing defectors. Especially, the isolated cooperator-driven punishers nearby the clusters or those at the tip of peninsulas locating at the borders of clusters are pioneers of expansions. The indispensable role of the two types of pioneers for CP clusters is revealed by the illustrated spectrum in Fig.~\ref{fig:rounddcp} in which one can find that whether the punishers owning less than three connected punishers have higher payoffs than those defective neighbors with the same neighborhood state is mainly responsible for the final dominance of punishers. As a result, there are considerable long-standing edges of corresponding states for optimal values of $A$ (the red patterns shown in middle plot of middle panels in Fig.~\ref{fig:rounddcp}). The similar mechanism leading to the dominance of defector-driven punishers in case D+DP is also revealed by the illustrations in Fig.~\ref{fig:roundddp} listed in Appendix. B. Whatever, both types of punishers are inclined to clustering because of their prosocial nature i.e., adopting cooperation strategy before exerting punishment. However, larger punishment fine is required as a remedy to weak network reciprocity caused by defector-driven punishers.

The observed behaviors in Figs.~\ref{fig:rounddcp} and \ref{fig:roundddp} provide a refined physical picture of clustering behaviors of punishers at two distinctly different stages: (1) At pre-cluster stage, owing to different sources to dive punishment executions, cooperator-driven punishers prefer to reduce punishment so as to preserve more competitive payoffs while defector-driven punishers have limited payoffs resulting from more frequent executions. Meanwhile support from network reciprocity is lacking, because compact clusters are not yet formed so early in the process. It turns out that cooperator-driven punishers are more likely to persevere to organize themselves into clusters, and therefore the required minimum size for growth of clusters is smaller than that of defector-driven punishers. (2) At the post-cluster stage, defector-driven punishers become instead conservative in punishing defectors. At this point, cooperator-driven punishers have strong support from the formed clusters on the one hand, and on the other hand there are enough sources to drive them to provide a sufficiently effective punishment. To sum up, in the two-strategy cases, cooperator-driven punishers are superior to defector-driven individuals in terms of both taking advantages of network reciprocity and suppressing defectors.

\begin{figure*}
\includegraphics[width=\linewidth]{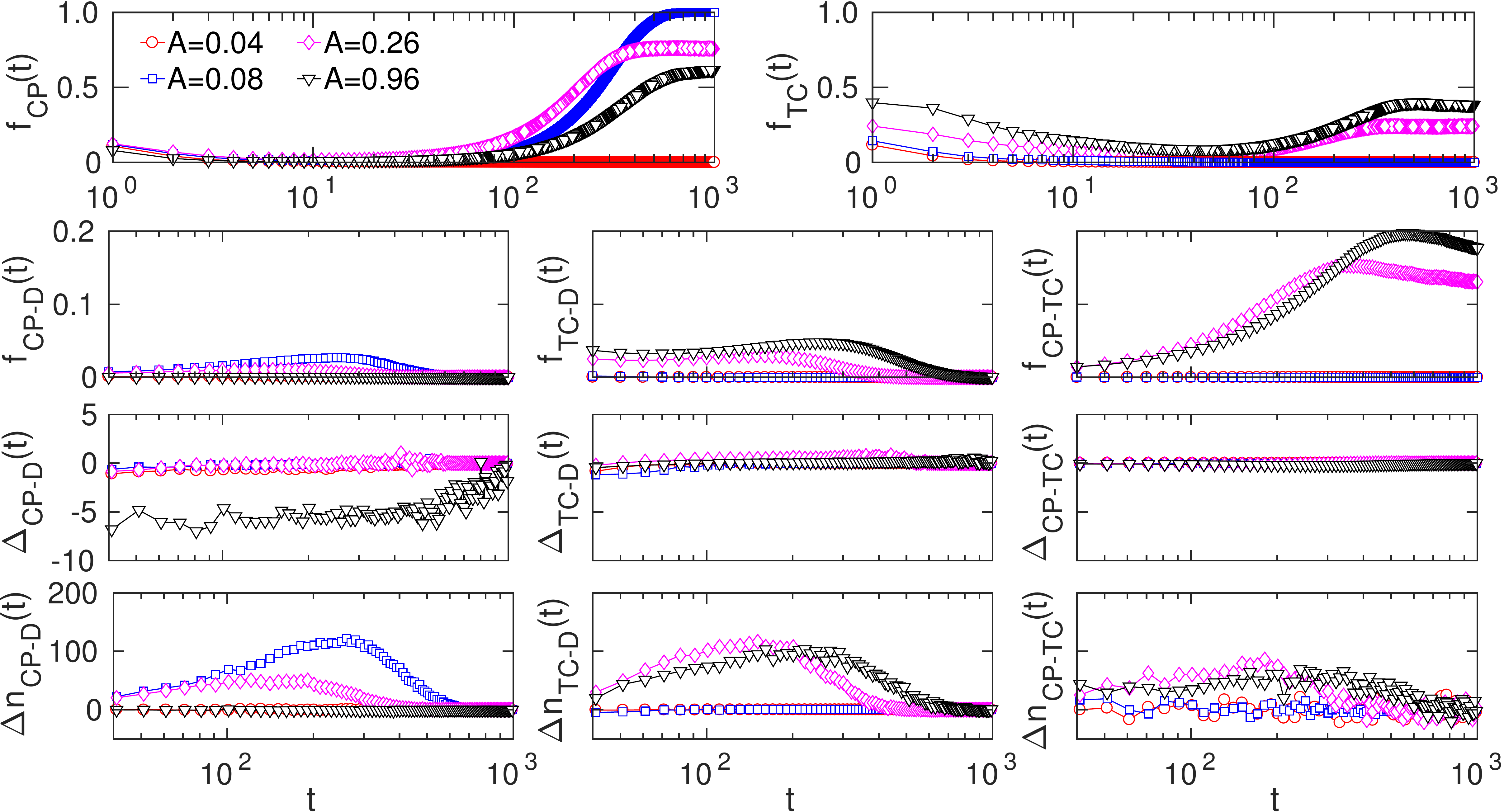}
\vspace{0cm}
\caption{The dynamic changes of statistical characterizing quantities for four different representative values of $A$ in  networked populations. In detail, from top to bottom rows, these statistical characterizing quantities are: the fractions of cooperator-driven punishers $f_{CP}(t)$ (traditional cooperators, $f_{TC}(t)$) in the population; the fractions of three different edges (CP-D, TC-D and CP-TC) which are normalized by the total number of edges in the population (the middle panels), where $f_{CP-D}(t)$, $f_{TC-D}(t)$ and $f_{CP-TC}(t)$  denote the fractions of corresponding edges respectively; the mean payoff gaps between the two ends of an edge at which an imitation process among different strategies happens, which is defined in the same manner proposed in caption of Fig.~\ref{fig:rounddcp}; the net increase of cooperator-driven punsihers (traditional cooperators) who are produced from the imitation process between defectors (traditional cooperators) and cooperator-driven punishers or traditional cooperators (cooperator-driven punishers) at time $t$ (the bottom panels), e.g. $\Delta n_{CP-D}(t)=n_{D\rightarrow CP}(t)-n_{CP\rightarrow D}(t)$ where $n_{D\rightarrow CP}(t)$ ($n_{CP\rightarrow D}(t)$) indicates the number of occurrences of imitation process which successfully translates a defector (cooperator-driven punisher) into a cooperator-driven punisher (defector) at time $t$. Other parameters are $r=4.0$ and $\alpha=4.0$.
}
\label{fig:rounddccp}
\end{figure*}

Fig.~\ref{fig:rounddccp} gives the quantitative traits of representative spatial evolution of the three competing strategies: TC, D and CP, for parameter values that yield different absorbing phases. Remarkably the additional participation of traditional cooperators can produce somewhat different evolution picture in which cooperator-driven punishers can possibly outperform defectors to leave survival spaces for traditional cooperators, and further form a stable coexistence with them. When $A$ is small, unresponsive cooperator-driven are naturally defeated by defectors and finally extincted, along with disappearance of TC. As the sensitivity $A$ increases, cooperator-driven punishers begin to conquer the whole network attributing to the same mechanism which has been uncovered in the two-strategy cases. This rises to occurrence of positive peak of $f_{CP-D}(t)$ (the fractions of edge CP-D) as well as $\Delta n_{CP-D}(t)$ (net increase of cooperator-driven punishers) (see Fig.~\ref{fig:rounddccp}). As the sensitivity is increased further, it can be observed slightly positive $f_{CP-D}(t)$ and larger positive peaks of $f_{TC-D}(t)$ indicated by pink markers in Fig.~\ref{fig:rounddccp}. In such cases, we have uncovered a strong mutualism between TC and CP, single-strategy clusters of whom cannot be persisted in the sea of defectors. Also, traditional cooperators at the borders of CP clusters actually play a role of 'protective film' which spatially isolate the punishers bearing additional punishment cost from those defectors, and thus prevent them from being eroded. This in turn gives these traditional cooperators i.e., second-order free riders advantaged position (slightly positive $\Delta_{TC-D}$ indicated by pink and black markers in Fig.~\ref{fig:rounddccp}) to outperform their defective neighbors whose payoffs have been greatly reduced by punishers within the same game groups. That is why $f_{TC-D}(t)$ and $f_{CP-TC}(t)$ are obviously positive while $f_{CP-D}(t)$ is approximated to zero (the second-row panels in Fig.~\ref{fig:rounddccp}). Meanwhile, the dynamics is formulated by the majority-like rule in the areas far from the borders of the clusters, on account of the fact that there is no difference between TC and CP individuals in absence of defectors. Correspondingly, there are a large number of edges CP-TC of which the two ends have equal payoffs ($\Delta_{CP-TC}(t)$ is approximately zero from beginning to end); however, positive $\Delta n_{CP-TC}(t)$ reveals a considerable translations from TC to CP (the forth-row panels in Fig.~\ref{fig:rounddccp}). 

The system also presents the similar dynamical traits in the case TC+D+DP, as shown in Figs.~\ref{fig:spatialdcdpfa},~\ref{fig:rounddcdp} listed in Appendix B. However, it is worth noting that DP clusters are not as strong as CP clusters in terms of resisting defectors, owing to the fact that defector-driven punishers become unresponsive as they clustering (as punishment-driven sources, defectors within the same game groups become fewer in number).  As another consequence, DP clusters are less favorable for the survival of surrounding traditional cooperators, and these cooperators are more dependent on network reciprocity so as to form more rounded clusters (the forth and fifth row panels in Fig.~\ref{fig:rounddcdp}). Thinner layers consisting of small TC clusters can also be found (see Fig.~\ref{fig:rounddcdp}).

We have checked that the results for the evolutionary situation D+CP+DP are consistent with the illustrations in Fig.~\ref{fig:functionanofour}(e1)(e2) that cooperator-driven punishers are prior to defector-driven ones. The results reveal that cooperator-driven punishers' prevalence depends mainly on that they are more successful in the battle against defectors. Furthermore, cooperator-driven punishers seem more essential for survival of defector-driven punishers. Since DP clusters cannot persist, and they have to combine with CP clusters who can fully take advantage of network reciprocity (for more details, see Fig.~\ref{fig:rounddcpdp}, along with corresponding descriptions).

\begin{figure*}
\includegraphics[width=\linewidth]{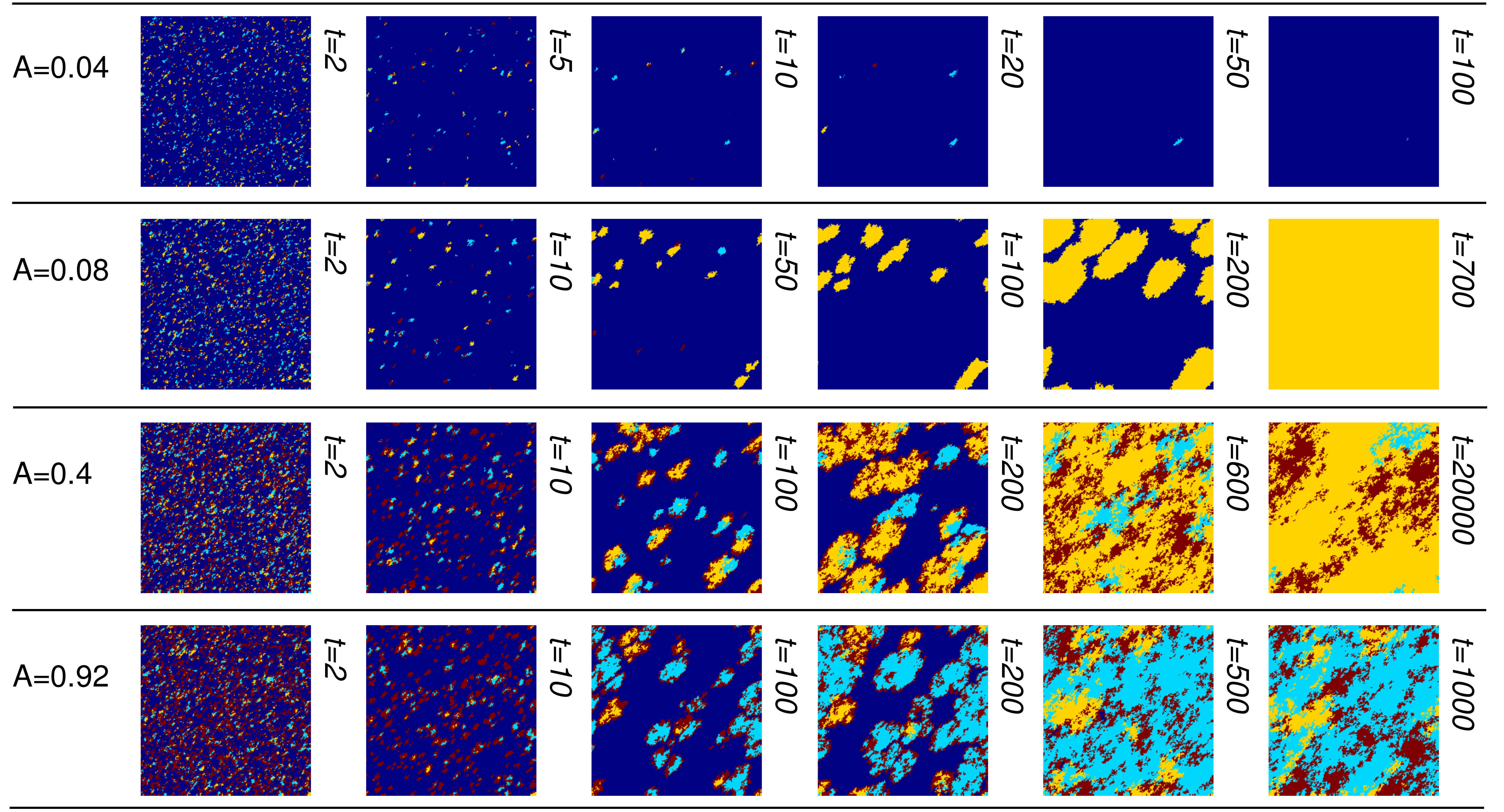}
\vspace{0cm}
\caption{Representative spatial evolution of the four strategies TC, D CP and DP in networked populations for four representative values of $A$. Depicted are snapshots of the Hexagonal lattice with size $L=200$, where the punishment fine is $\alpha=4.0$. Cooperators (defectors) are shown in dark red (dark blue), while cooperator-driven (defector-driven) punishers are depicted in orange (light blue).
}
\label{fig:spatialcdcpdpfa}
\end{figure*}

Given the condition that all four strategies are present, one can expect the following evolutionary patterns of different strategy clusters shown in Fig.~\ref{fig:spatialcdcpdpfa}, based on the mechanisms uncovered for three-strategy cases: Unlike TC and DP clusters, CP clusters can exist alone in the face of defectors; while DP or TC clusters have to combine with each other or with CP clusters. DP or CP clusters are more or less enfolded by traditional cooperators for large sensitivity $A$, indicating the establishment of an additional mutualism between traditional cooperators and the two types of punishers. The majority-like rule still formulate the dynamics in the interiors of nondefectors' clusters where there is no essential distinction between punishers and traditional cooperators, which can be considered as the key mechanism behind the competitive relationship between the two types of punishers, as also shown in FIg.~\ref{fig:functionanofour}. CP is superior to TC and DP for a wide range of the sensitivity $A$ especially when the parameter is intermediate. Furthermore, Figs.~\ref{fig:functionanofour}  reveals that the above-mentioned microscopic mechanisms for three-strategy cases do not rely on any additional strategic complexity limiting its general validity. 

\begin{figure*}
\includegraphics[width=\textwidth]{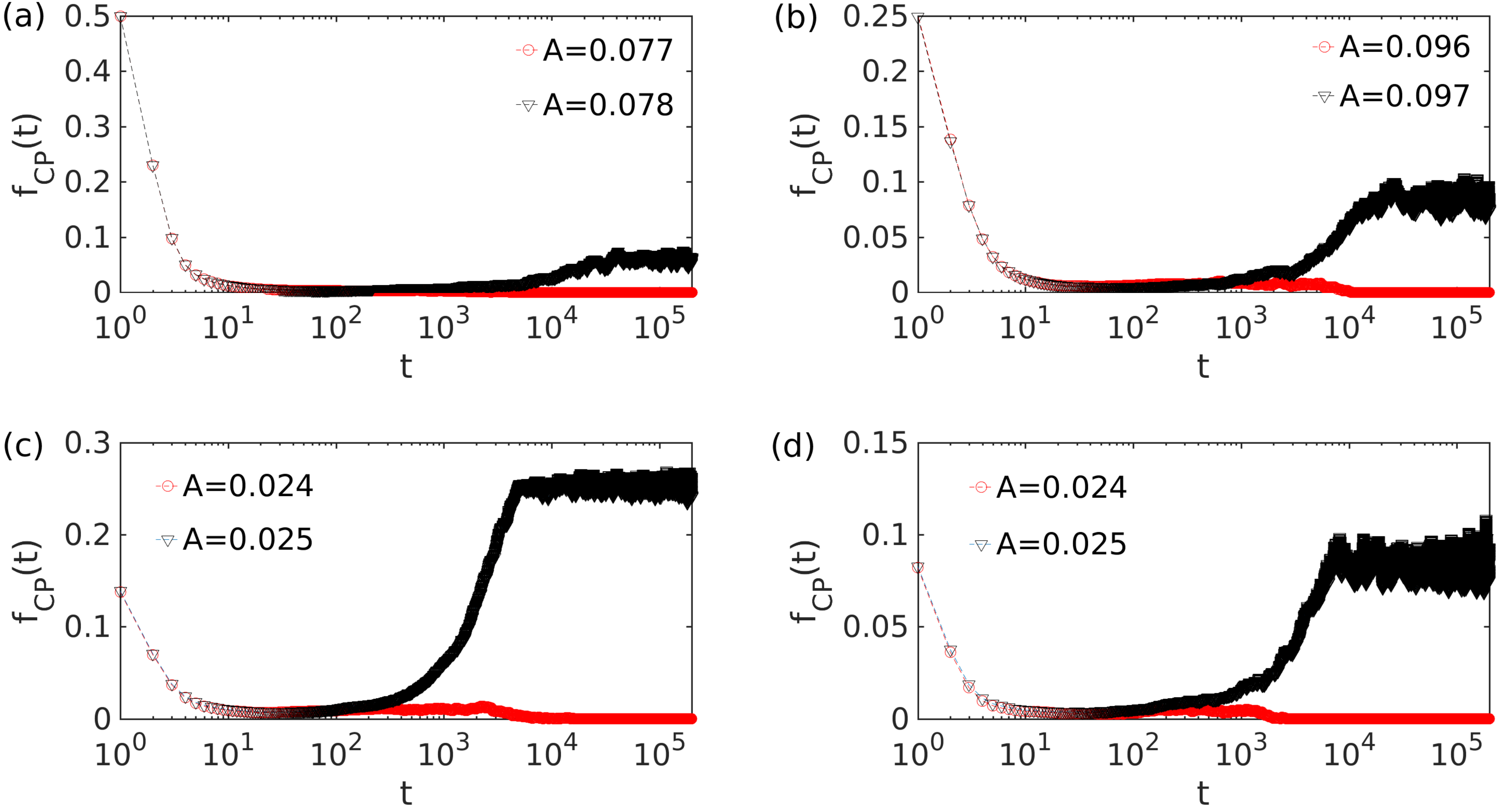}
\caption{Evolution of the fractions of cooperator-driven punishers over time, starting with a random initial state. The results are presented for four different evolutionary situations: (a) D+CP, (b) TC+D+DP, (c) D+CP+DP and (d) TC+D+CP+DP, and in the close vicinity of transition points. The values of punishment fine are (a) $\alpha=1.8$, (b) $\alpha=4.4$, (b) $\alpha=6.0$ and (d) $\alpha=6.0$, respectively. The other parameter is $r=4.0$.
}
\label{fig:MC_data}
\end{figure*}

Since stochastic imitation of a neighboring strategy ensures the existence of (homogeneous) absorbing states and critical phase transitions. We have also explicitly explored the dynamic behaviors of the system in the close vicinity of transition points above which the system shifts from FD phase to FP phase. In the case of D+CP, it can be observed in Fig.~\ref{fig:MC_data}(a) that in the close vicinity of transition point, the final outcome in the two cases is remarkably different while the difference in the sensitivity $A$ is minute, which is a characteristic feature of discontinuous phase transition. Moreover, similar but more obvious discontinuous behaviors in terms of the fractions of either DPs or CPs can be produced in the cases TC+D+DP, D+CP+DP and TC+D+CP+DP (see Fig.~\ref{fig:MC_data}(b)-(d)) and other evolutionary situations. This is in accordance with what have been uncovered from the previous studies involving pool punishment against defectors~\cite{perc2012}.

\begin{figure*}
\includegraphics[width=\linewidth]{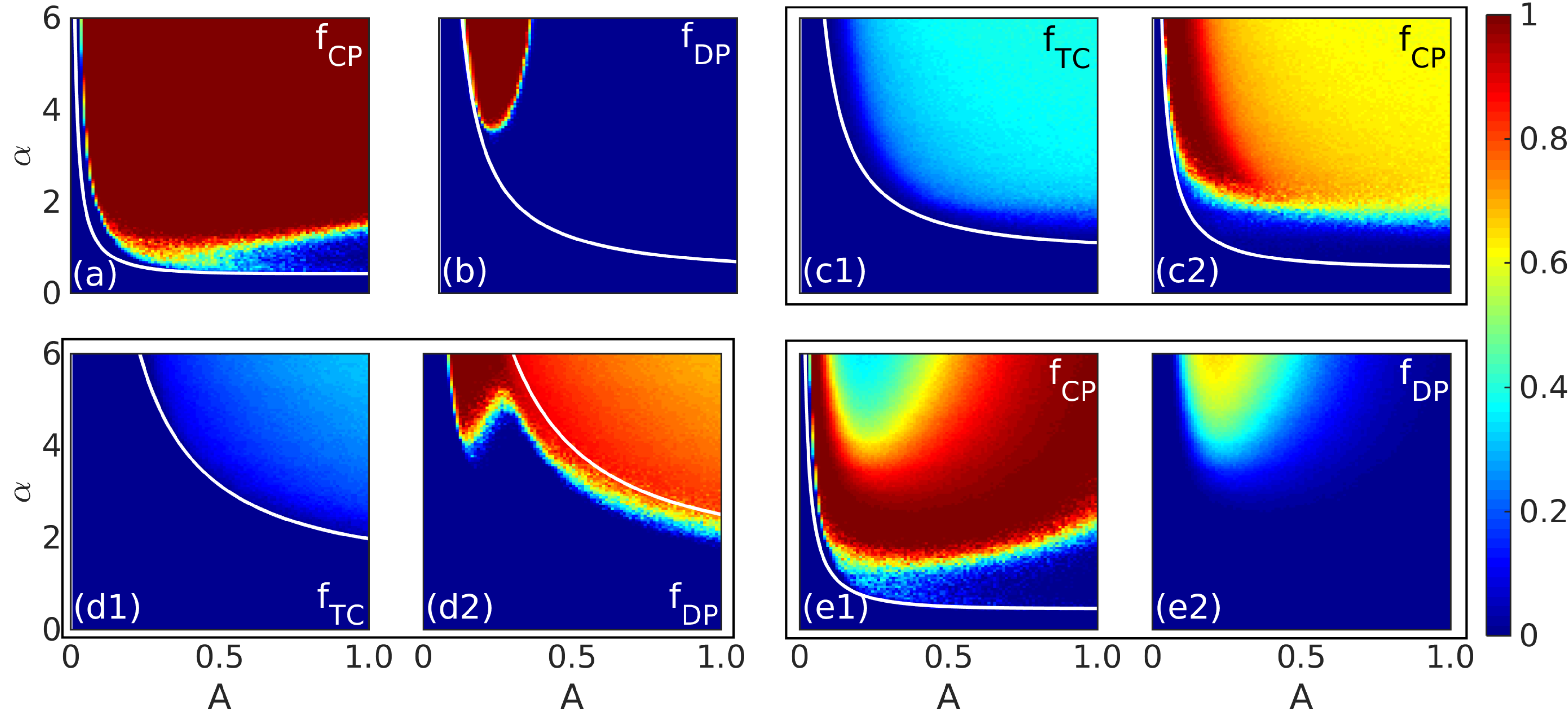}
\vspace{0cm}
\caption{The dependence of simulated final steady fractions of different nondefection strategies on both $A$ and $\alpha$, where the results for five different evolution situations are respectively illustrated: (a) D+CP, (b) D+DP, (c1) and (c2) TC+D+CP, (d1) and (d2) TC+D+DP, (e1) and (e2) D+CP+DP. The population is networked, and the other parameter is $r=4.0$. Correspondingly, the value of $f_{s}$ used to semi-analytically estimate the boundary lines are (see Appendix~C for further details): (a) $f_{CP}=1.0$, (b) $f_{DP}=0.96$, (c1) $f_{TC}=0.1$ and $f_{CP}=0.55$, (c2) $f_{TC}=0.1$ and $f_{CP}=0.71$ (d1) $f_{TC}=0.115$ and $f_{DP}=0.465$, (e1) $f_{CP}=0.1$ and $f_{DP}=0.84$.
}
\label{fig:simulationnofour}
\end{figure*}

\begin{figure*}
\includegraphics[width=\linewidth]{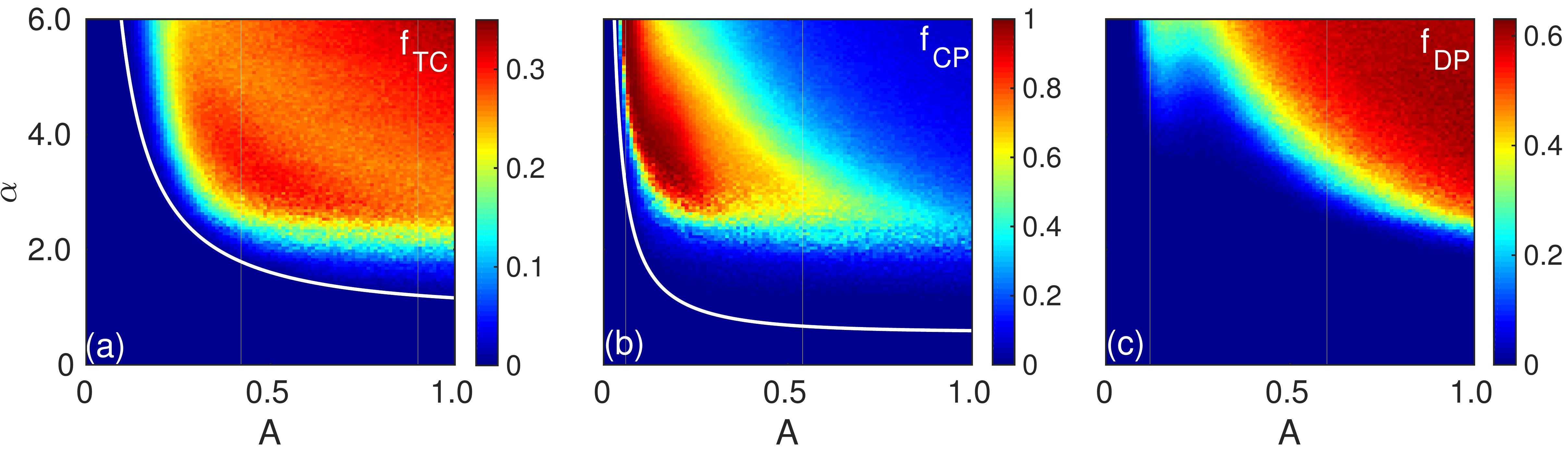}
\vspace{0cm}
\caption{The dependence of simulated final steady fractions of three different nondefection strategies on both $A$ and $\alpha$. The population is networked, where four strategies TC, D, CP and DP are considered. Correspondingly, the values of $f_{s}$ used to semi-analytically estimate the boundary line are: (a) $f_{TC}=0.09$, $f_{DP}=0.09$ and $f_{CP}=0.46$; (b) $f_{TC}=0$, $f_{DP}=0$ and $f_{CP}=0.8$. The other parameter is $r=4.0$.
}
\label{fig:simulationfour}
\end{figure*}

Finally, Figs.~\ref{fig:simulationnofour} and \ref{fig:simulationfour} provide a comprehensive picture in the parameter plane of ($A$, $\alpha$) of the evolutionary dynamics, as well as semi-analytically estimated boundary lines (see Appendix~C for more details). The boundaries between the regions of different nondefective strategies are not given. By means of the displayed results in Fig.~\ref{fig:MC_data}, note that a discontinuous phase transition always occurs when the system shifts from FD phase to FP phase, instead of continuous phase transition from FD phase to SP phase; which is irrespective of the complexity of evolutionary situations. The discontinuous phase transition is due to the positive feedback arising from the fact that moderately sensitive punishers begin to firstly clustering, and then keep growing to completely dominate the whole population along with the extinction of traditional cooperators; while the continuous phase transition stems from increasing competitiveness of punishers strengthen by punishment $\alpha$~\cite{szolnoki2017} and persistence of traditional cooperators.  Overall, CP can more effectively enforce a regulation to promote and sustain public goods through enlarging the regions of ND phase, and thus be more prompt than DP; especially in absence of traditional cooperators. More specifically, performances of cooperator-driven punishers can be weakened by free-riding behaviors of traditional cooperators (see Fig.~\ref{fig:simulationnofour}(a) and \ref{fig:simulationnofour}(c1), \ref{fig:simulationnofour}(c2)) who, however, may greatly help defector-driven punishers to beat defectors in a much larger parameter space (see Fig.~\ref{fig:simulationnofour}(b) and \ref{fig:simulationnofour}(d1), \ref{fig:simulationnofour}(d2)). In both cases, two types of punishers in turn provides survival space for traditional cooperators, and hence regions of TC more or less coincide the those of DP or CP phase (see Fig.~\ref{fig:simulationnofour}(c1),\ref{fig:simulationnofour}(c2) and \ref{fig:simulationnofour}(d1), \ref{fig:simulationnofour}(d2)), but not totally. In the presence of more than two strategies, the optimal parameter regions for different nondefective strategies repel each other especially between the two punishing strategies, as a result of the majority-like rule. Meanwhile, this rule rises to another notable result that performance of CP is more or less limited by DP. Also, nonmonotonic changes of freqwe alternatively uencies of both punishing strategies are more clearly illustrated, which is the consequence of both network reciprocity and the majority-like rule among nondefective individuals. Finally, notice that the semi-analytic estimated boundaries successfully distinguish the simulated regions of NP phases from the whole parameter space, although there is deviation in the case of D+DP for large $A$ (Fig.~\ref{fig:simulationfour}(b)). This discrepancy is attributed to a vicious circle between defectors and defector-driven punishers: more defectors the punishers punish, lower payoffs they have, and thus more of them get eroded. Finally, there are more defectors to further drive the punishers to exert punishments with a higher probability. Consequently, defector-driven punishers go extinct in a much larger parameter region than theoretical expectation based on well-mixed situation; which is exacerbated in the ranges of large $A$. 

By comparing with the results shown in Figs.~\ref{fig:newconditionnofour} and \ref{fig:simulationfour} with corresponding analytical predictions given by Figs.~\ref{fig:analysisnofour} and \ref{fig:ratioanalysisfour} with the initial conditions unchanged, we can conclude that networked structure of the population is essential for superior performance of cooperator-driven punishers to defector-driven ones. Hence the system can achieve more desirable level of public goods.  

\section{Discussions and Conclusions}
\label{sec:discuss}

To summarize, empirical explorations of corporate self-regulation and government regulation suggest two different punishment measures which play indispensable roles in regulatory issues, calling for a general game model to take into account both the types of punishments. We have accomplished that in this paper. For the purpose of fully identifying both the interactions among different strategies and the performances of the two punishments, we have considered six different evolutionary situations with and without traditional cooperation, in which one or both punishment strategies are introduced to fight against defectors. In addition, we have proposed a theoretical approach to  completely describing the evolutionary dynamics of the six different combinations of strategies in an infinite well-mixed population. Next, agent-based simulations are employed to give numerical results for networked populations embedded on a regular lattice. At the same time, we have  developed a semi-analytical method which allows us to give relatively accurate estimations of the boundaries between full-defection and nondefection phases in most evolutionary situations.

For the infinite well-mixed population, we firstly use a series of replicator equations capturing features of present model to give gradients of selection for two-strategy cases, phase portraits for three-strategy cases, and ratios of attraction basins of nondefection for the cases with more than two strategies. The study of the population has revealed that involvement of both punishments is more effective than one of the punishments alone in sustaining cooperation, based on two respects: larger scopes of the attraction basins of full nondefection and larger region sizes of nondefection phases. The mechanism behind this result can be attributed to the fact that abundant available punishments against defectors are provided by the two types of punishers at different evolution stages, respectively. Next, the analytical results have suggested monotonous effect of synergy effect, punishment fine and feedback sensitivity on facilitating the advantages of nondefectors in terms of scopes of the attraction basins of full nondefection. Further support about the same effect is obtained by giving a comprehensive picture of strategy fractions in the parameter plane of $(A,~\alpha)$ for six different evolution situations: When both $A$ and $\alpha$ are large enough, nondefection phases appears. Since both frequent available punishments and large cost of one punishment are imposed on defectors. Of particular note is that cooperator-driven punishment is overall uncompetitive in presence of defector-driven punishment, however, slightly more favorable for nondefectors. In addition, punishment fine and feedback sensitivity turn out to be the key parameters to govern the performances of punishment measures in such population. By means of the semi-analytical method, we have given boundaries to accurately distinguish nondefection phases from full defection phases for each evolutionary situation, through roughly estimating the fractions of different nondefective strategies nearby the boundaries. Whatever, traditional cooperators can undermine the evolutionary advantages of punishers~\cite{panchanathan2004,Sigmund2010,perc2012}, even though in the desirable situations that the cooperation is further promoted by both the punishment measures together.

For networked population, agent-based simulations of the evolutionary dynamics generates more rich results. In comparison to the findings under infinite well-mixed condition, networked structure are overall more favorable for the survival or even dominance of nondefectors i.e. sustaining the public cooperation, which is supported by the comparison of comprehensive pictures for two different populations. We have obtained a physical original of this phenomenon through a detailed statistical analysis of the emerging spatial patterns in terms of the frequencies of different types of strategies and edges, mean payoff gaps between the two ends of edges connecting two individuals with different strategies, net increase of different nondefectors, payoff-gap and number spectrums of different states of edges. The analysis has revealed that it can be attributed to two major factors: (1) support from network reciprocity in any case; (2) mutualism betweeen traditional cooperators and punishers that works in the cases with traditional cooperation. Quite remarkably, the punishers help traditional cooperators to reduce the competitive ability of defectors in the vicinity of punishers' clusters to some extent. Whereas traditional cooperators who are effectively second-order free-riding on the punishments, form an active layer around punishers, which protects them against defectors. Mutualism between the punishers and traditional cooperators is thus established. Especially as a consequence of the strong mutualism, traditional cooperators can counterintuitively and largely facilitate the prevalence of defector-driven punishers. While it turns out that cooperator-driven punishers are always vulnerable to the same second-order free-riding. Another interesting point is that cooperator-driven punishment is proved to be a more powerful measure than defector-driven punishment with respect to enlarging the scopes of favorable parameters and promoting cooperation, by virtue of their great ability to fully take advantage of network reciprocity. This finding is in accordance with the empirical conclusions that self-regulation is superior to government regulation for benefiting consumers, businesses and the economy~\cite{Gunningham1997,Short2010,Lenox2003}.

Moreover, unlike what happens in the infinite well-mixed population, to have a desirable evolutionary outcome with high-level cooperation in the network, an intermediate range of feedback sensitivity is surprisingly needed. While monotonous effect of synergy effect, punishment fine and feedback sensitivity on facilitating the advantages of traditional cooperators rather than punishers has been still identified. The statistical analysis of spatial pattern formations has also provided a physical understanding. For small sensitivity, punishments from unresponsive punishers are lacking, and thus unable to sustain nondefectors's survival. Conversely, if the sensitivity is large, over-sensitive punishers would punish too many defectors so that they cannot have competitive payoffs in comparison to defectors. Therefore intermediate sensitivity is an optimal choice of punsihers. We have found that under such parameter condition punishers cannot only defeat defectors through sufficient punishments but also maintain competitive advantages to get clustering in time, leading to persistent growth of nondefectors' clusters. Furthermore, in the vicinity of the borders of nondefectors' clusters, isolated cooperator-driven punishers or those at the tip of peninsulas are found to be pioneers of expansions. In addition, it is also worth noticing that our semi-analytical approach fails to give a relatively accurate boundary in the situation with defectors and defector-driven punishers, given that feedback sensitivity is large. This discrepancy is attributed to the vicious circle arised from strategic nature of defector-driven punishers, which also causes the poorer performance of defector-driven punishment in networked population. Finally, explorations of the frequencies of different strategies as function of feedback sensitivity or both the sensitivity and punishment fine have disclosed a competitive relationship among nondefectors, especially among cooperator-driven and defector-driven punsihers. This is the result of the majority-like rule which frequently happens within nondefectors' clusters.     

Relating to the reality, we conclude our work by providing two general remarks. Firstly, our study has uncovered potential favorable conditions for operations of self-regulation and government regulation. More precisely, in the social or economic systems with imperfect information induced by spatial structure, self-regulation can be accepted as a useful tool to sustain commons and eliminate the conflicts of interest; particularly when self-regulation organizations have an intermediate response speed, and whose regulations are strong enough. Conversely, if the state of the whole system is known to the individuals (like internet system), mix of the two types of regulations may be a better choice to achieve optimal public goals, such as internet co-regulation scheme~\cite{Marsden2011}. From another perspective, our study has given a possible interpretation of why self-regulation has recently begun to gain wide acceptance and interest for applications~\cite{Gunningham1997,Langevoort2002,Krawiec2003,Lenox2003,Estlund2010}: imperfect information or spatial limitation in the markets or social systems. Secondly, we can conclude from our study that regulating effects from response speed of regulation organizations are highly dependent upon information transparency (i.e. structure) of the systems. More precisely, high information transparency let high response speed be always essential for a good running market. In contrast, when information transparency is rather low because of spatial limitation, selecting an intermediate response speed is a better choice for regulation organizations such as SROs. 

Finally, we must stress that our present model does capture top-down prescriptive rules of government regulation, as well as third party certification schemes or government whatchdogs~\cite{thirdorganization,Marsden2011} which may result in better firm behavior. For self-regulation, we avoid the adversarial problem about putting the fox in charge of the hen house; which is beyond our present research. Whatever, our present study has developed a computational and theoretical paradigm to understand the relative roles played by SROs and government regulation (external powerful force such as troops) in the framework of game theory, which has potential implications not only to self-regulation but also to other topics in economics and political science. We hope to be able to extend our analysis to temporal networks~\cite{Holme2012} or multilayer networks~\cite{Gomez2012,Wang2013,Kivela2014,Wang2015}, as well as to more complex situations by considering above mentioned realistic mechanisms or antisocial punishment~\cite{herrmann2008,Rand2010,horne2016,szolnoki2017}.

\section*{Acknowledgments}
This work was partially supported by China Postdoctoral Science Foundation (PSF) under Grant No. 2015M582532 and by the National Natural Science Foundation of China under Grants No. 61433014, No. 61503062, No. 11575072.

\newpage
\section*{Appendix A}
\label{sec:appendixa}
The evolutionary dynamics of the studied system can be analytically described by a set of replicator equations~\cite{weibull1997,Huang2018}:
\begin{eqnarray}
\frac{df_{s}(t)}{dt}=f_{s}(t)(\Pi_{s}-\overline{\Pi})
\label{eq:finalreplicator}
\end{eqnarray}, where $f_{s}$ and $\Pi_{s}$ indicate the fractions of individuals owning strategy $s\in \{TC,~D,~CP,~DP\}$ and their correspondingly expected payoff in theoretical analysis, respectively. While $\overline{\Pi}=\sum\limits_{s}f_{s}(t)\Pi_{s}$ denotes the average payoff of the entire population. Theoretically the expected payoff for each strategy $s$ could be further given by
\begin{eqnarray}
\Pi_{s}=\sum\limits_{0\leq N_{i}\leq G-1}\frac{(G-1)!}{\prod\limits_{i}N_{i}!}\prod\limits_{i}f_{i}^{N_{i}}\Pi^{'}_{i}.
\label{eq:finalpay}
\end{eqnarray}
Furthermore, taking the case D+CP as an example, $\Pi^{'}_{i}$ can be obtained according to the following method:
\begin{widetext}
\begin{eqnarray}
\label{eq:paydcp1}
\Pi^{'}_{D}  & = & (1.0-g_{CP})^{N_{CP}}(\frac{rN_{CP}}{G})+[1.0-(1.0-g_{CP})^{N_{CP}}](\frac{rN_{CP}}{G}-\alpha), \quad g_{CP}=\frac{AN_{CP}}{G};\\
\label{eq:paydcp2}
\Pi^{'}_{CP}  & = & (1.0-g_{CP})(\frac{r(N_{CP}+1)}{G}-1.0)+g_{CP}[\frac{r(N_{CP}+1)}{G}-1.0-\sum\limits_{i=0}^{N_{CP}}\frac{N_{CP}!}{i!(N_{CP}-i)!}g_{CP}^{i}(1.0-g_{CP})^{(N_{CP}-i)}\frac{N_{D}\alpha}{i+1}], \notag \\
 \quad g_{CP} & = & \frac{A(N_{CP}+1)}{G}. 
\end{eqnarray}
\end{widetext}
We can easily extend the above method to the other five evolutionary situations.

\section*{Appendix B}
\label{sec:appendixb}

\begin{figure*}
\includegraphics[width=\textwidth]{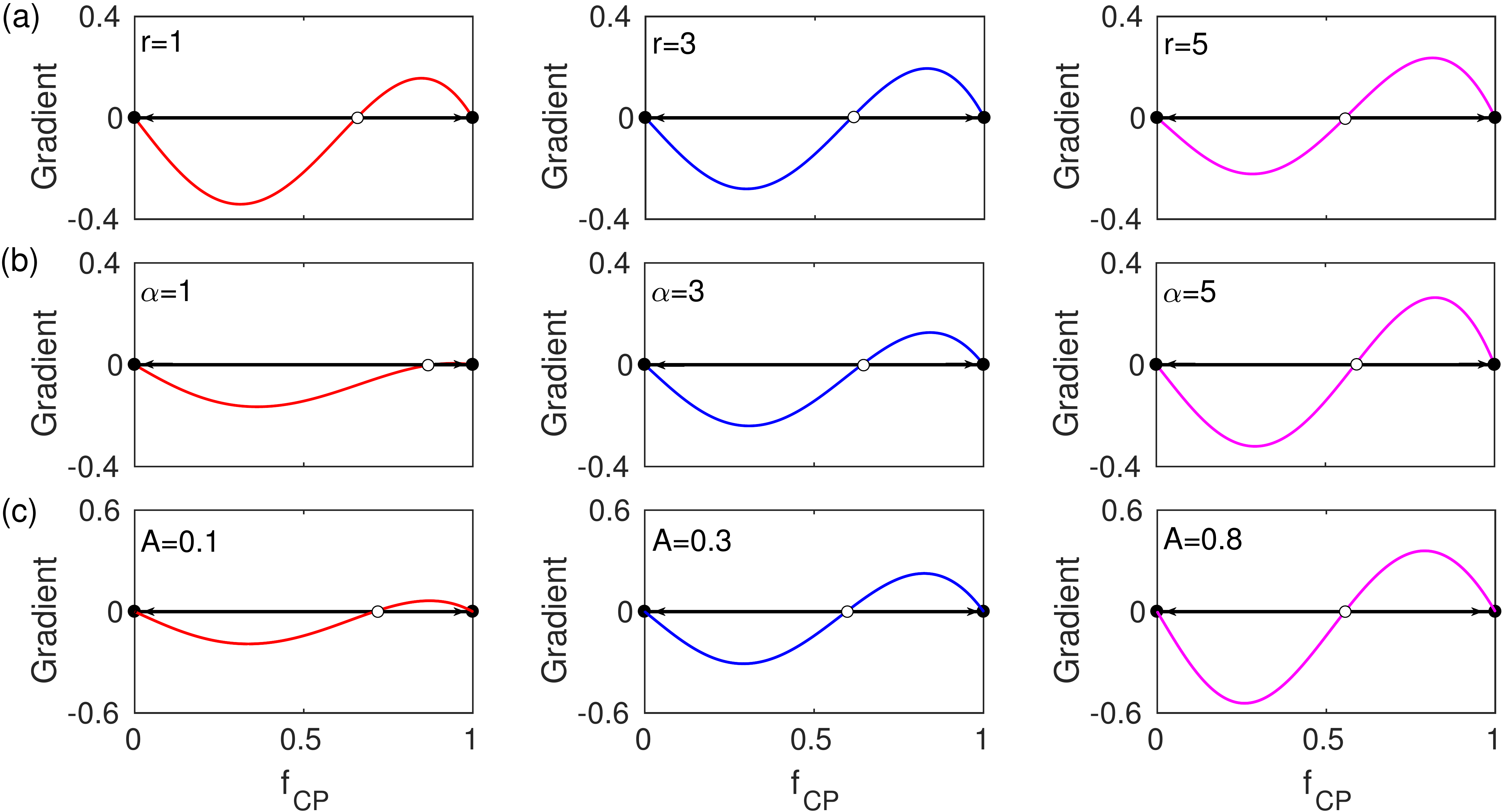}
\vspace{0cm}
\caption{Gradient of selection in dependence on the fraction of cooperator-driven punishers for evolutionary situation D+CP. Stable steady states $f_{CP}= 0$ and $f_{CP}= 1$ are depicted with solid circles, while the unstable steady state is depicted with an open circle. Arrows indicate the expected direction of evolution. The arrow pointing to the right indicates that cooperator-driven punishment is favored over defection. Panel (a) shows results for three different values of $r$ with $A = 0.25$ and $\alpha=4.0$; (b) shows results for three different values of $\alpha$ with $A = 0.25$ and $r=3.0$; (c) shows results for three different values of $A$ with $\alpha = 4.0$ and $r=3.0$. In any case, the intermediate state is unstable on condition that left side of the gradient is negative while the right is positive. 
}
\label{fig:gtwodcp}
\end{figure*}
Behaviors of the selection gradient $df_{CP}/dt$ for different parameter conditions are presented in Fig.~\ref{fig:gtwodcp}. The illustrations show that CP can transform the defined game into a coordination game with full CP and full D (FD) as the two stable equilibria, along with an intermediate unstable steady state (i.e., coexistence state of the two strategies). By the means of the rule that larger gradient indicates higher speeds at which the system converges to the stable equilibrator, we can state that FD is more attractive than full CP, regardless that large $r$ ($\beta$ and $A$) can help full CP to be more advantaged to some extent.  

\begin{figure*}
\includegraphics[width=\textwidth]{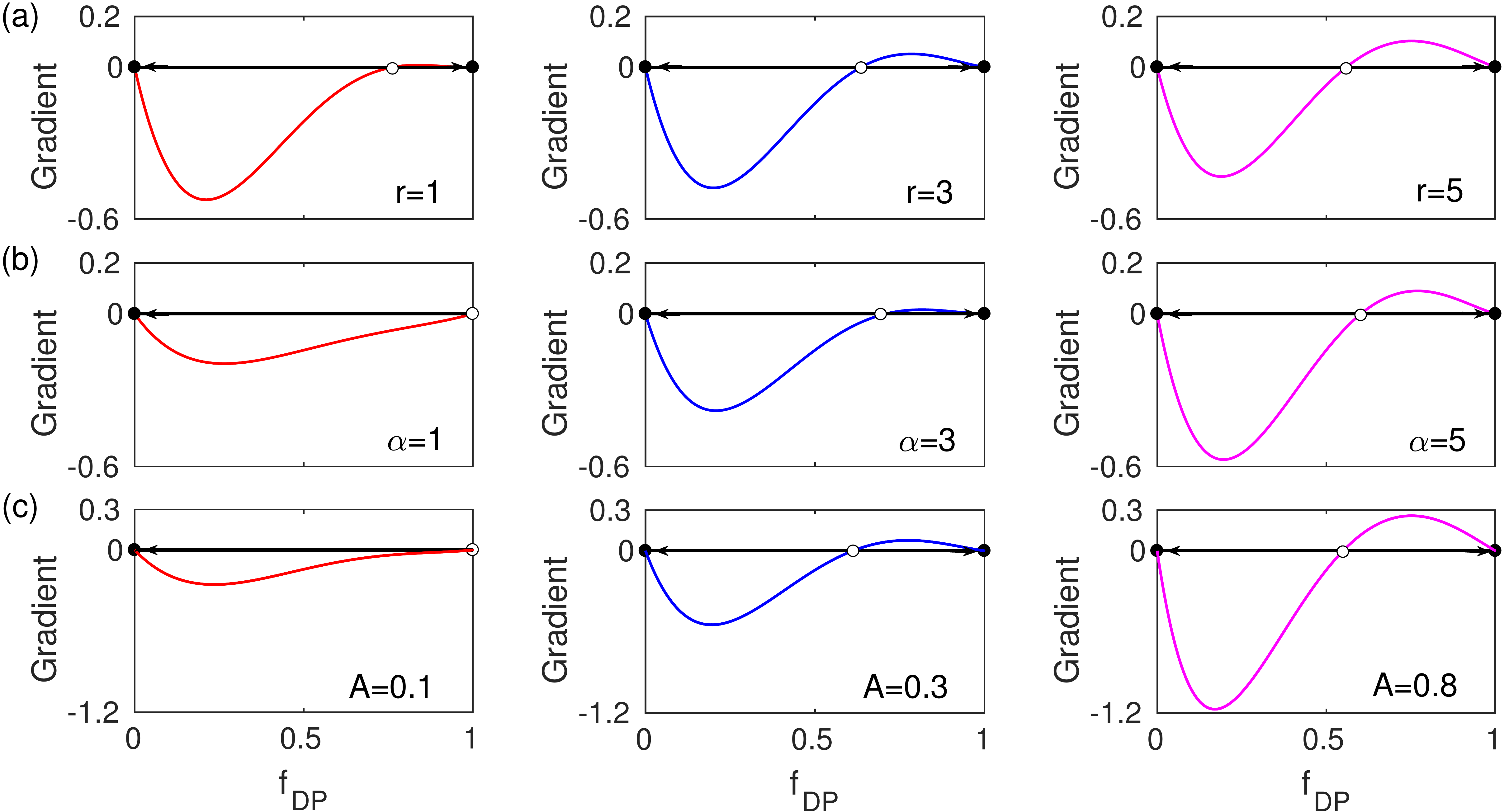}
\vspace{0cm}
\caption{Gradient of selection in dependence on the fraction of defector-driven punishers for evolutionary situation D+DP. In most cases,  $f_{DP}= 0$ and $f_{DP}= 1$ are stable steady states, while the coexistence state of the two strategies are unstable or even impossible. Likewise, arrows indicate the expected direction of evolution, and DP is favored over D if the arrow points to the right. Panel (a) shows results for three different values of $r$ with $A = 0.25$ and $\alpha=4.0$; (b) shows results for three different values of $\alpha$ with $A = 0.25$ and $\alpha=3.0$; (c) shows results for three different values of $A$ with $\alpha = 4.0$ and $r=3.0$.
}
\label{fig:gtwoddp}
\end{figure*}

Likewise, as shown in Fig.~\ref{fig:gtwoddp}, DP can still transform the defined game into a coordination game with full DP and FD as the two stable equilibria except the first panel listed in Fig.~\ref{fig:gtwoddp}(b) and (c). We can also find that the system can more quickly reach the state of FD than that of full DP. Figs.~\ref{fig:gtwodcp} and \ref{fig:gtwoddp} provide a key hint that the position of the coexistence state can be used to measure how facilitated different parameters are for the punishment. More specifically, lower position value of the intermediate coexistence state is, more likely the system is to reach the sate of full punishment (FP).

\begin{figure}
\includegraphics[width=0.5\textwidth]{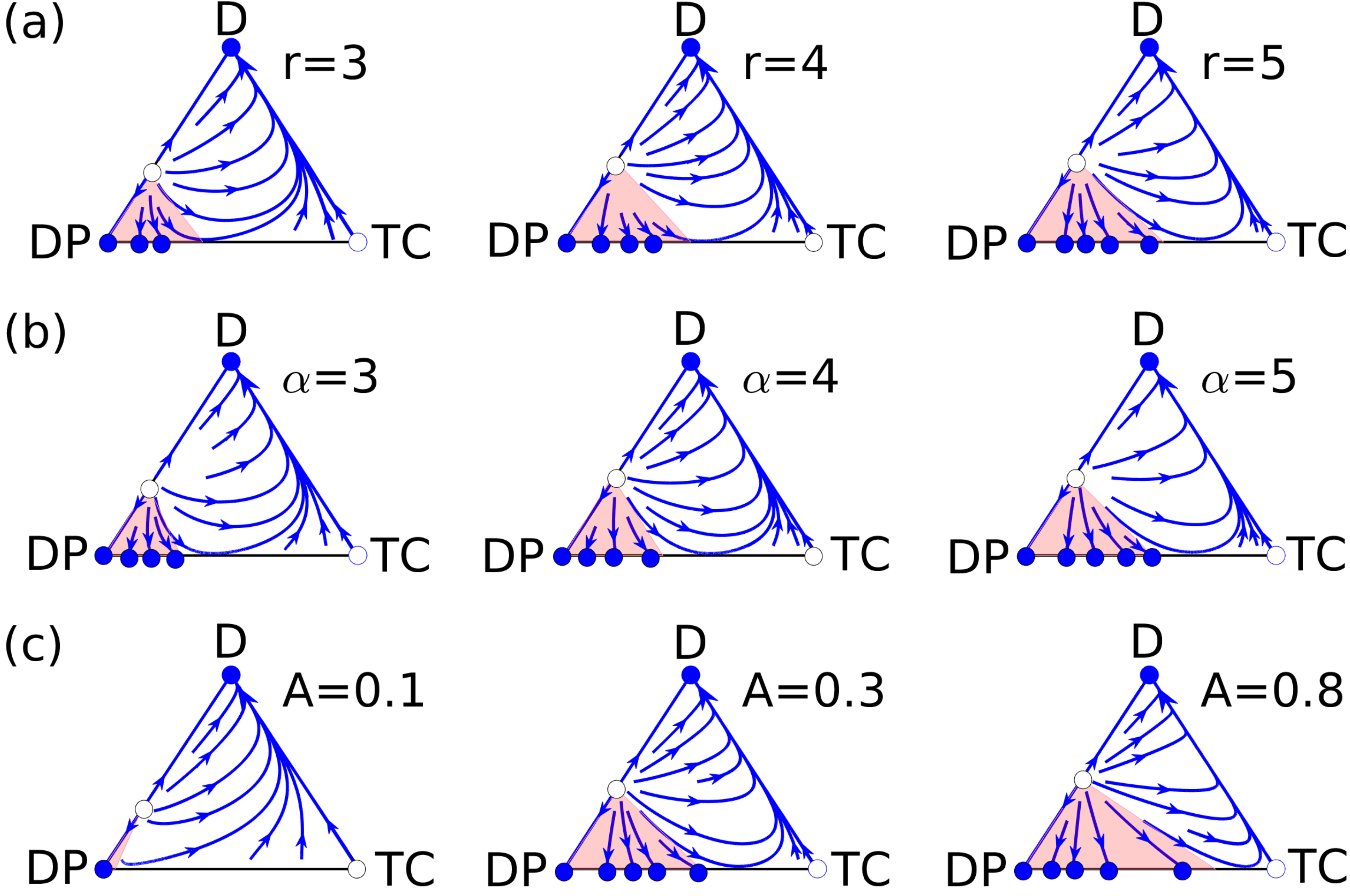}
\vspace{0cm}
\caption{The phase portraits of the system for the evolutionary situation: TC+D+DP. (a) The portraits for three representative values of synergy factor are illustrated, where the other parameters are $A=0.25$ and $\alpha=4.5$. (b) The portraits for three representative values of punishment fine are illustrated, where the other parameters are $A=0.25$ and $r=4.5$. (b) The portraits for three representative values of sensitivity are illustrated, where the other parameters are $\alpha=4.5$ and $r=4.5$. The solid triangle vertexes indicate attractors of the system, while the hollow one means a repellor. The red regions indicate attraction basins where the system converges to a state of either full DP or SP.
}
\label{fig:threecddp}
\end{figure}

Fig.~\ref{fig:threecddp} provides a comprehensive picture of the system dynamics for the case TC+D+DP, exhibiting rich phenomena. Depending on the initial conditions, the system will evolve towards one of the following three states: FD, stable coexistence of DP and TC (i.e., state of segment punishers, SP), and full DP; except the state of full cooperation (FC) or coexistence of the three strategies. It is obvious that the three strategies fail to form a cyclic dominance~\cite{szolnoki2017}. Note that attraction basin of nondefectoin (ND) gets larger with increasing $r$, $\alpha$ or $A$, which further confirms that the monotonous effects of the three parameters in promoting public goods is despite of the intervention of traditional cooperation. Specifically, achieving SP state largely depends on whether there are adequate initial defector-driven punishers or not, especially for large $r$, $\alpha$ or $A$. Another obvious feature is that defectors only have opportunity to completely conquer entire population after that defector-driven punishers have gone extinction (i.e., punishment is absent) resulting from exploitation of traditional cooperators, exhibiting that the evolution trajectories which ended in FD state have to firstly reach the side of simplex with two corners 'D' and 'TC'. 

\begin{figure}
\includegraphics[width=0.5\textwidth]{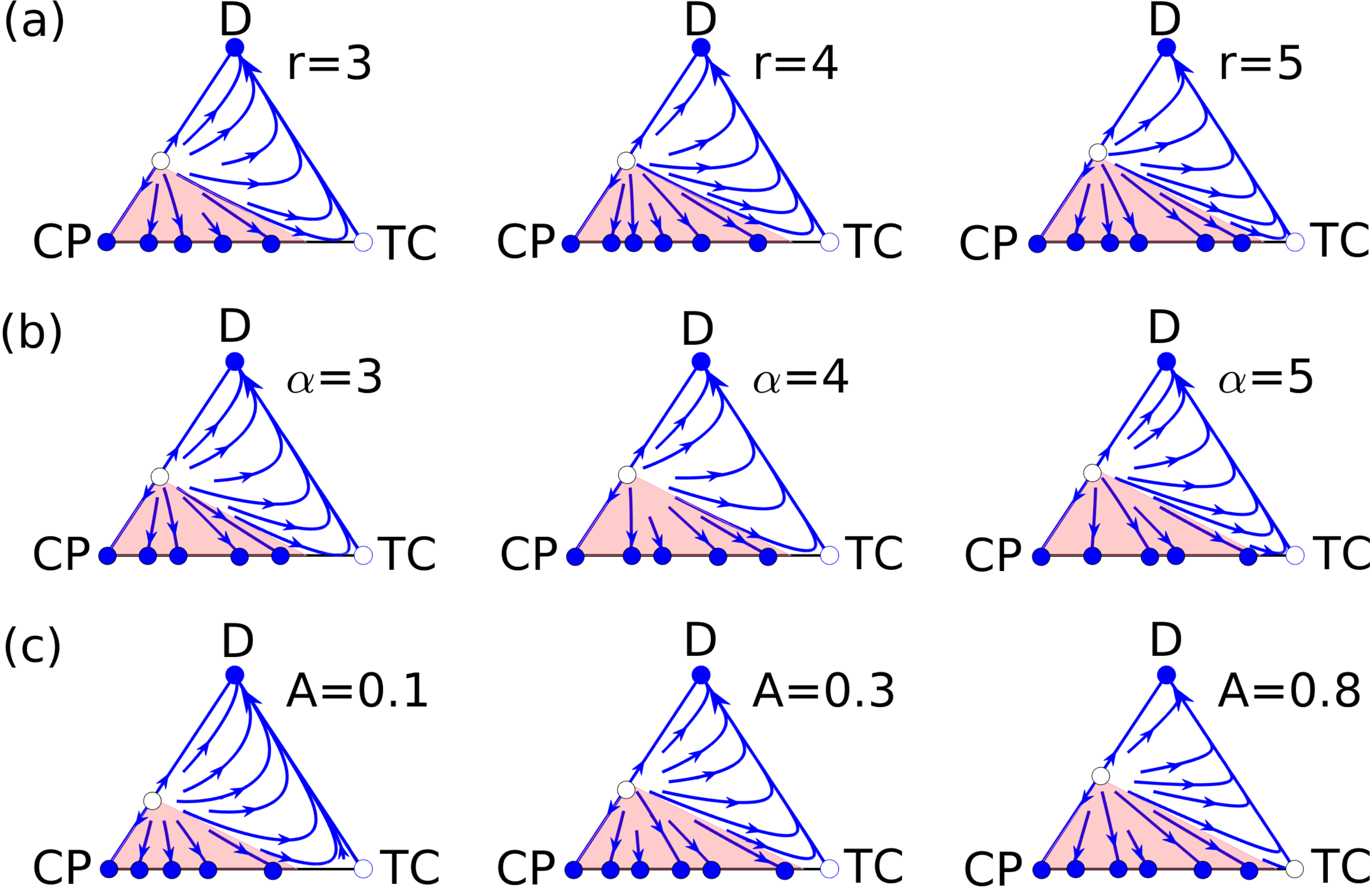}
\vspace{0cm}
\caption{The phase portraits of the system for the evolutionary situation: TC+D+CP. (a) The portraits for three representative values of synergy factor are illustrated, where the other parameters are $A=0.25$ and $\alpha=4.5$. (b) The portraits for three representative values of punishment fine are illustrated, where the other parameters are $A=0.25$ and $r=4.5$. (b) The portraits for three representative values of sensitivity are illustrated, where the other parameters are $\alpha=4.5$ and $r=4.5$. The solid triangle vertexes indicate attractors of the system, while the hollow one means a repellor. The red regions indicate attraction basins where the system converges to a state of either full CP or SP.
}
\label{fig:threecdcp}
\end{figure}

As shown in Fig.~\ref{fig:threecdcp}, three strategies TC, D and CP together generate the similar results especially with respect to both the patterns of attraction basins and evolution trajectories. The only difference is that CP performs better in facilitating more favorable initial conditions under which the system evolves towards SP state. Thus, combing with the results given by Fig.~\ref{fig:threecddp}, CP shows a slightly greater advantage than DP in sanctioning t[!h]hose defectors. Whatever, Figs.~\ref{fig:threecddp} and \ref{fig:threecdcp} show that the basins of attraction for SP is less than half of the simplex, which reveals that one of the two punishers alone fails to sustain cooperation alone unless the punishers initially capture the majority of the population. Since both the punishments are challenged by second-order free-riding from traditional cooperators.

\begin{figure}
\includegraphics[width=0.5\textwidth]{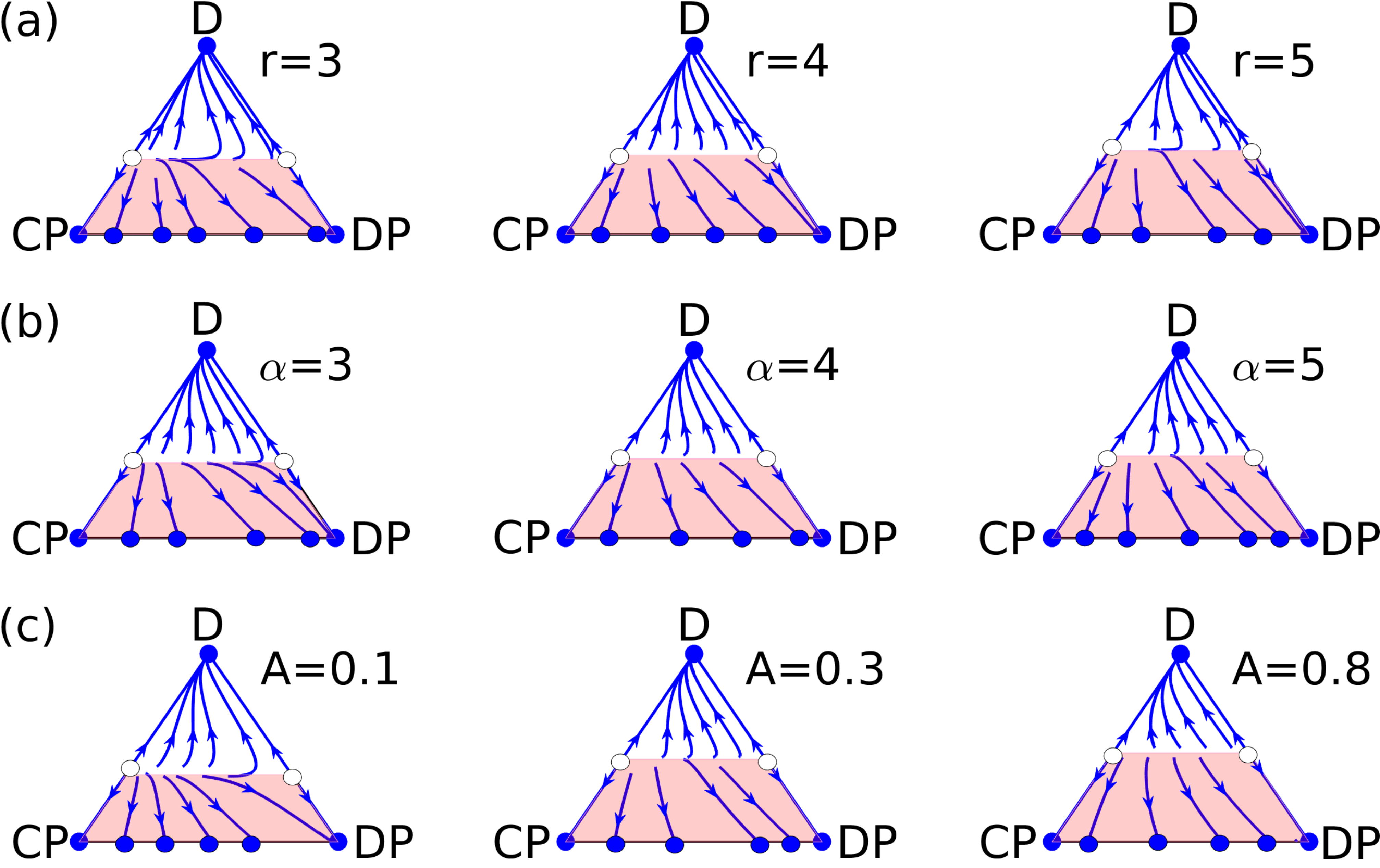}
\vspace{0cm}
\caption{The phase portraits of the system for the evolutionary situation: TC+CP+DP. (a) The portraits for three representative values of synergy factor are illustrated, where the other parameters are $A=0.25$ and $\alpha=4.5$. (b) The portraits for three representative values of punishment fine are illustrated, where the other parameters are $A=0.25$ and $r=4.5$. (b) The portraits for three representative values of sensitivity are illustrated, where the other parameters are $\alpha=4.5$ and $r=4.5$. Three solid triangle vertexes suggest that each of the three strategies is a attractor of the system. The red regions indicate attraction basins where the system converges to a state of FP (full CP or full DP) or SP.
}
\label{fig:threeccpdp}
\end{figure}

Fig.~\ref{fig:threeccpdp} provides a different picture in which cooperator-driven and defector-driven punishers together can effectively repel  defectors in most cases. More surprisingly, this phenomenon is robust to the changes of punishment fine, synergy factor and feedback sensitivity, as well as that increasing the three parameters can enhance competitive advantage of the two punishers to some extent. This suggests a non-trivial interplay between defector-driven and cooperato-driven punishment in promoting the public cooperation (including TC, CP and DP). Still, a stable interior point is still absent in such case. Based on the illustrations in Fig.~\ref{fig:threeccpdp}, we can say mix of the two types of punishments is better in preventing the entire population from being eroded by the first-order free riders. 

\begin{figure*}
\includegraphics[width=\linewidth]{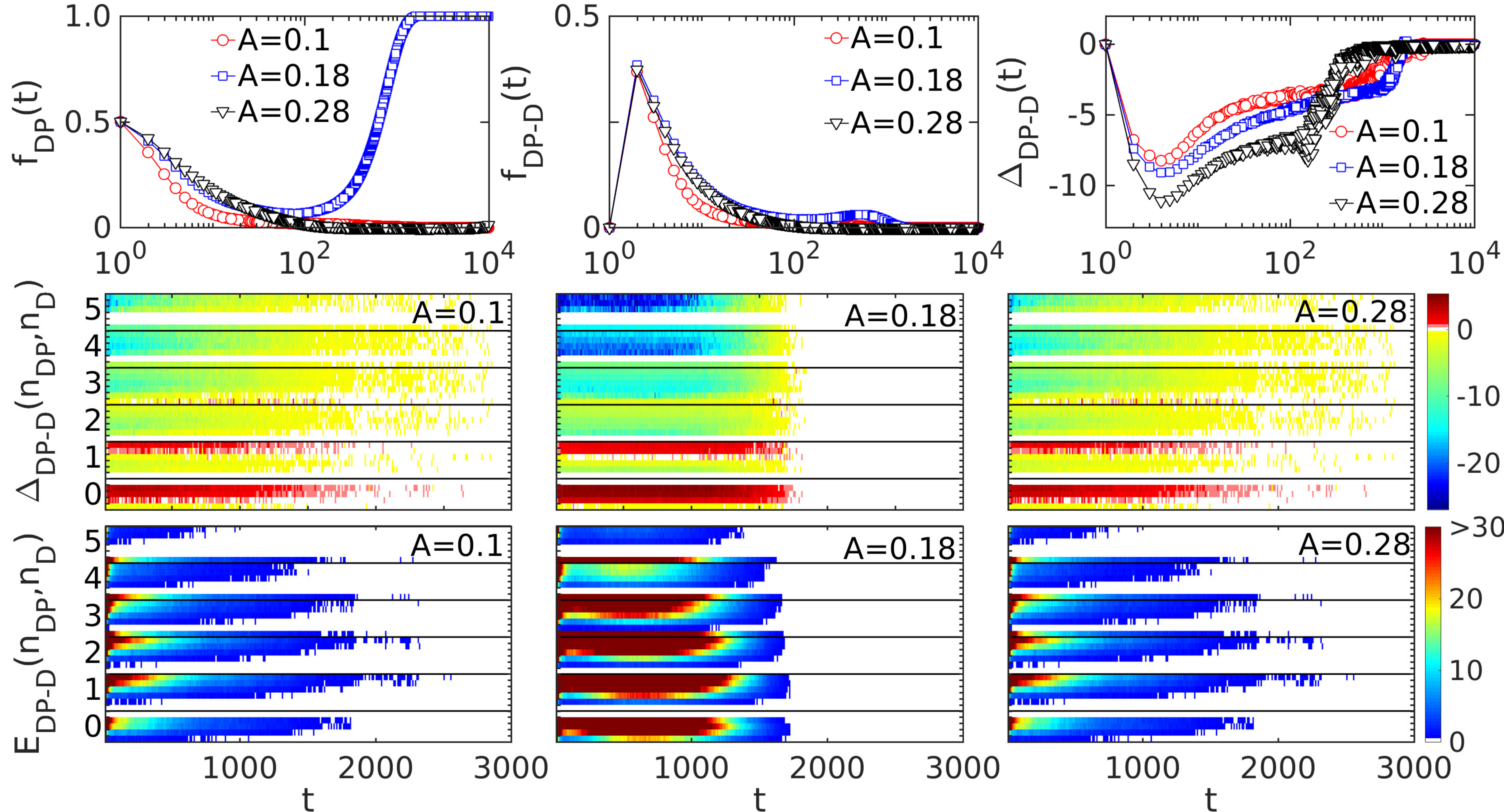}
\vspace{0cm}
\caption{Further illustrations of the evolutionary dynamics of the networked populations in case D+DP for three representative values of $A$. Shown are the behaviors of the five different statistical characterizing quantities. Other parameters are $r=4.0$ and $\alpha=4.0$.
}
\label{fig:roundddp}
\end{figure*}

Also, in a similar way, the growth of DP clusters start with individuals at the tips of peninsulas and nearby isolated ones, as shown in Figs.~\ref{fig:roundddp}. In the same way, small or large sensitivity induces uncompetitive uncompetitive defector-driven punishers. Intermediate sensitivity is optimal for the punishers.

\begin{figure*}
\includegraphics[width=\linewidth]{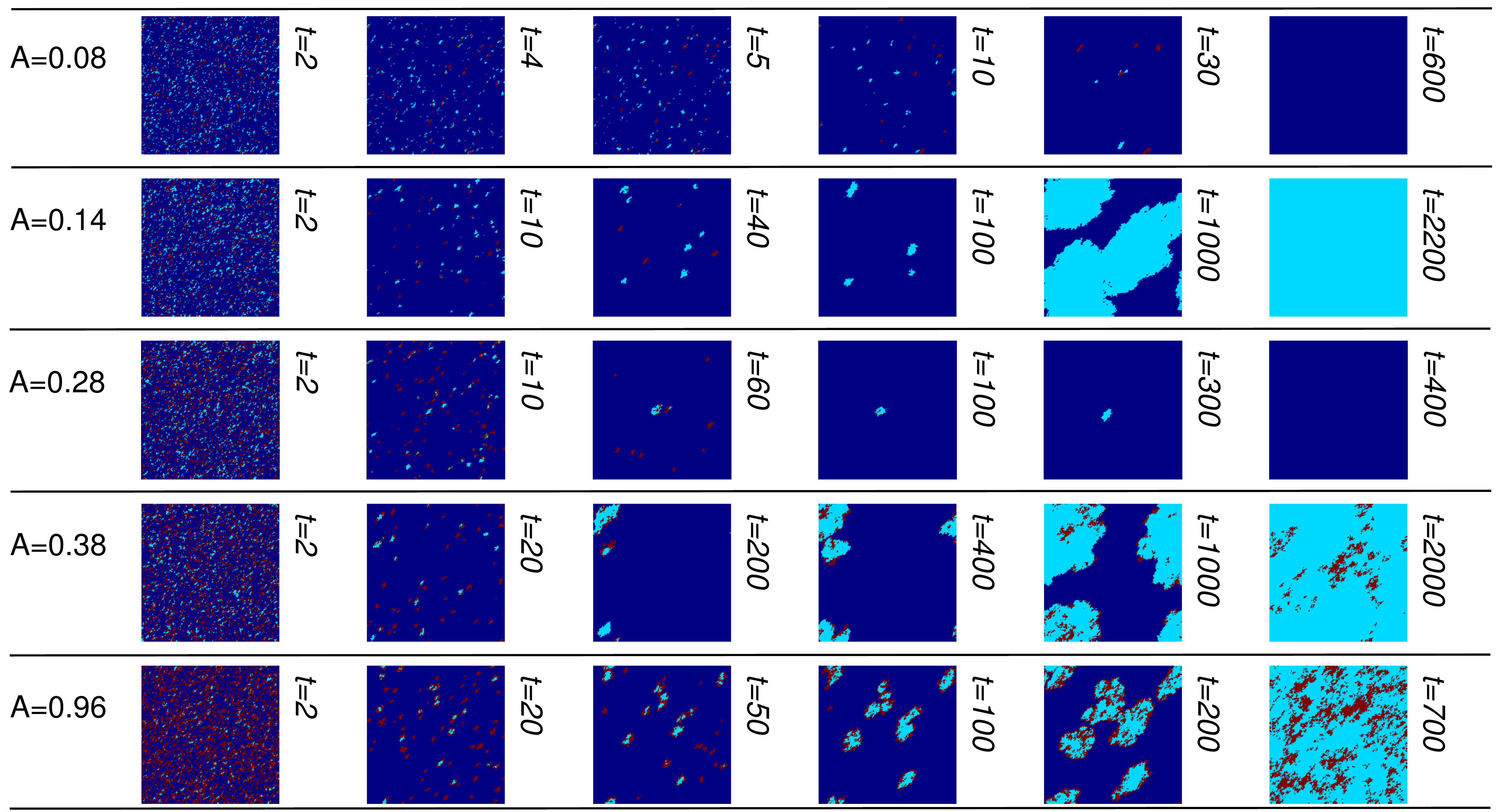}
\vspace{0cm}
\caption{In networked populations, spatial evolution of the three competing strategies TC, D and DP, for four representative values of $A$. Depicted are snapshots of the Hexagonal lattice with size $L=200$, where the punishment fine is $\alpha=4.4$. The color codes are the same as in Fig.~\ref{fig:spatialcdcpdpfa}.
}
\label{fig:spatialdcdpfa}
\end{figure*}

\begin{figure*}
\includegraphics[width=\linewidth]{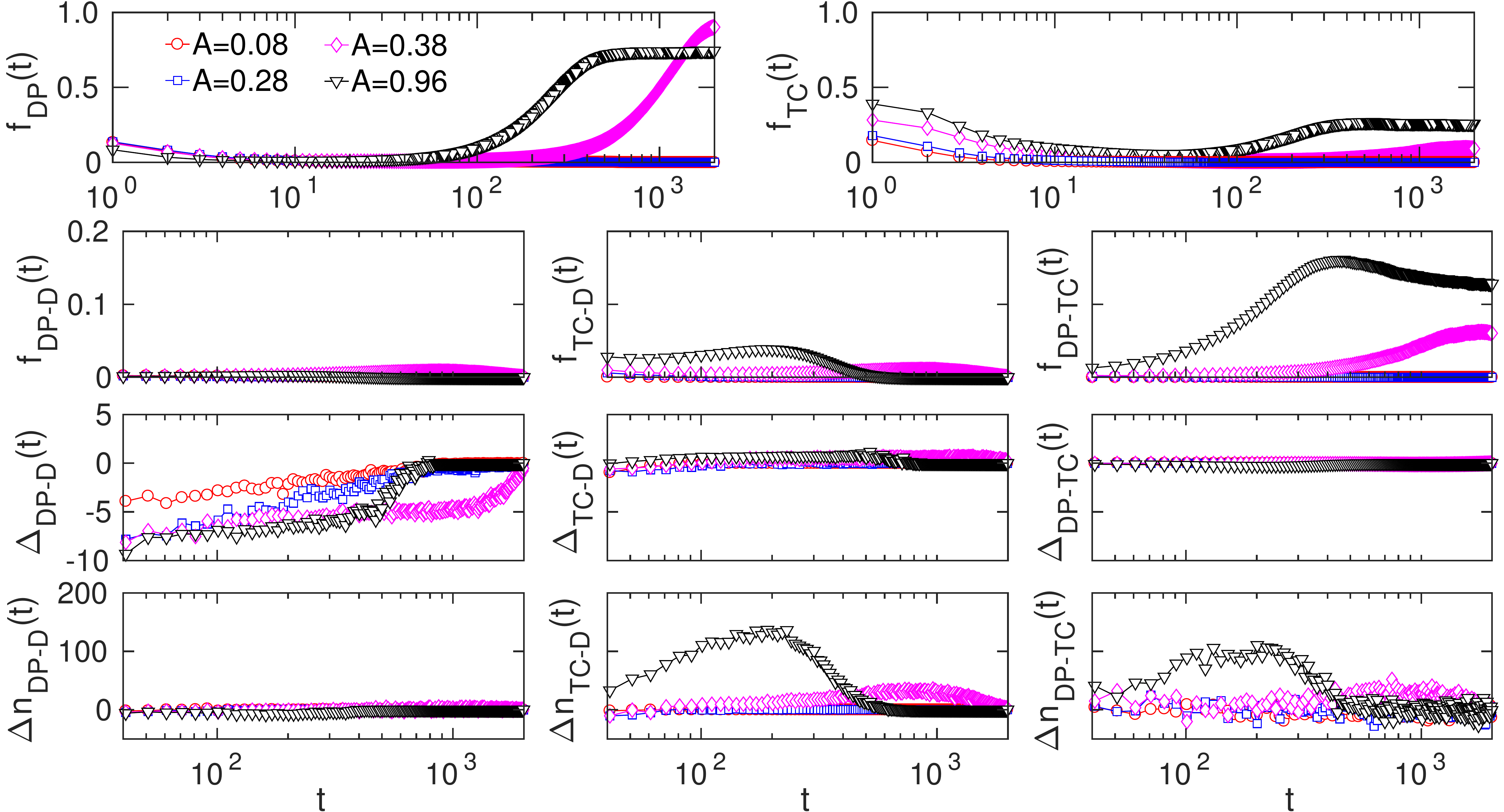}
\vspace{0cm}
\caption{Understanding of roles of different strategies in evolutionary process happening in networked populations in presence of three competing strategies TC, D and DP. Shown are the behaviors of three classes of  statistical characterizing quantities for four representative values of $A$ corresponding to the four groups of spatial snapshots presented in Fig.~\ref{fig:spatialdcdpfa}. The other parameters are $r=4.0$ and $\alpha=4.4$.
}
\label{fig:rounddcdp}
\end{figure*}

Figs.~\ref{fig:spatialdcdpfa} and \ref{fig:rounddcdp} further support that there exist an optimal intermediate range of feedback sensitivity to facilitate the complete dominance of defector-driven punishers in presence of traditional cooperators, which is also considered as a desirable evolutionary outcome with both first-order and second-order free-riding. As shown in Fig.~\ref{fig:spatialdcdpfa}, traditional cooperators at the borders of clusters still play a role of 'protective film' which spatially isolate the punishers bearing the punishment cost from those defectors. Meanwhile this leaves these traditional cooperators a chance to beat their defective neighbors whose payoffs have been greatly reduced by DP. Compared with the case TC+D+CP, spatial pattern formations shown in Figs.~\ref{fig:spatialdcdpfa} indicate similar trajectory of evolution. However, except in the case of large intermediate feedback sensitivity punishers are too active to lost the territorial battle with defectors, further leading to extinction of all nondefectors. In the same way, there is a strong strong mutualism between defector-driven punishers and traditional cooperators.

\begin{figure*}
\includegraphics[width=\linewidth]{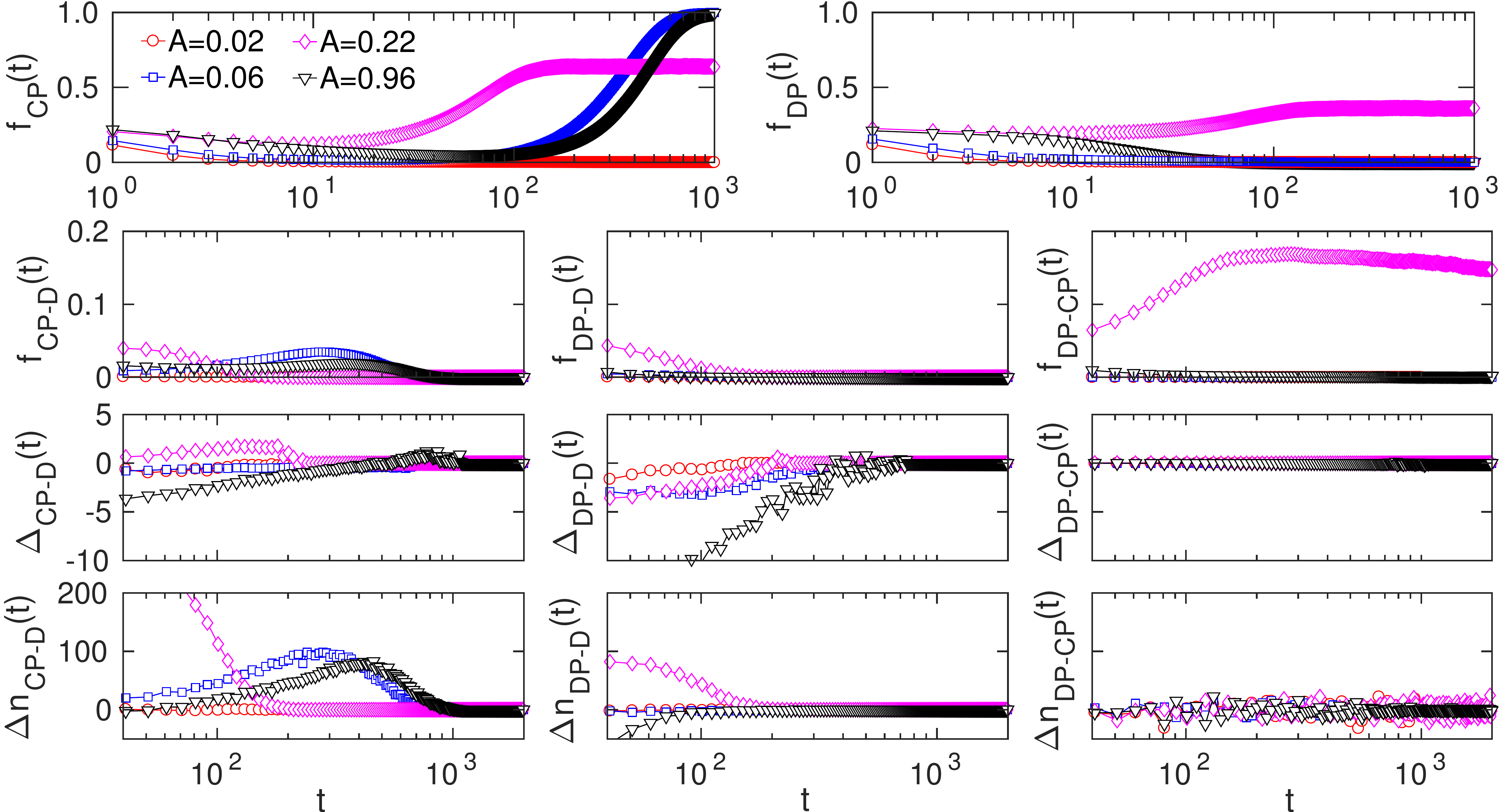}
\vspace{0cm}
\caption{Understanding of roles of different strategies in evolutionary process happening in networked populations, where three strategies D, CP and DP are considered. Shown are the behaviors of three classes of statistical characterizing quantities for four representative values of $A$. The other parameters are $r=4.0$ and $\alpha=4.0$.
}
\label{fig:rounddcpdp}
\end{figure*}

The results for the evolutionary situation D+CP+DP are presented in Fig.~\ref{fig:rounddcpdp}, enabling a direct comparison between the two types of punishers. The illustrations are consistent with the results in Fig.~\ref{fig:functionanofour}(e1)(e2) that cooperator-driven punishers are prior to defector-driven punishers. The illustrations defector-driven ones. Especially positive peaks of both $\Delta_{CP-D}(t)$ and $\Delta n_{CP-D}(t)$ are found to be larger than those of both $\Delta_{DP-D}(t)$ and $\Delta n_{DP-D}(t)$ for the same parameter setting, while $\Delta n_{DP-CP}(t)$ are always approximated to zero. The phenomena reveal cooperator-driven punishers' prevalence depends mainly on that they are more successful in the battle against defectors. $\Delta_{CP-D}(t)$ ($\Delta n_{CP-D}(t)$) is thus larger than $\Delta_{DP-D}(t)$ ($\Delta n_{DP-D}(t)$) (see the third and forth rows of panels in Fig.\ref{fig:rounddcpdp}). However, in the areas without defectors their competition still frequently happens in the form of majority-like rule, leading to fluctuations of $\Delta n_{DP-CP}(t)$ exhibited in Fig.~\ref{fig:rounddcpdp} and further exacerbating the divide between the two types of punishers. At the same time, cooperator-driven punishers seem more essential for survival of defector-driven punishers. Since DPs cannot persist alone, and they have to combine with CPs who can fully take advantage of network reciprocity.

\begin{figure*}
\includegraphics[width=\linewidth]{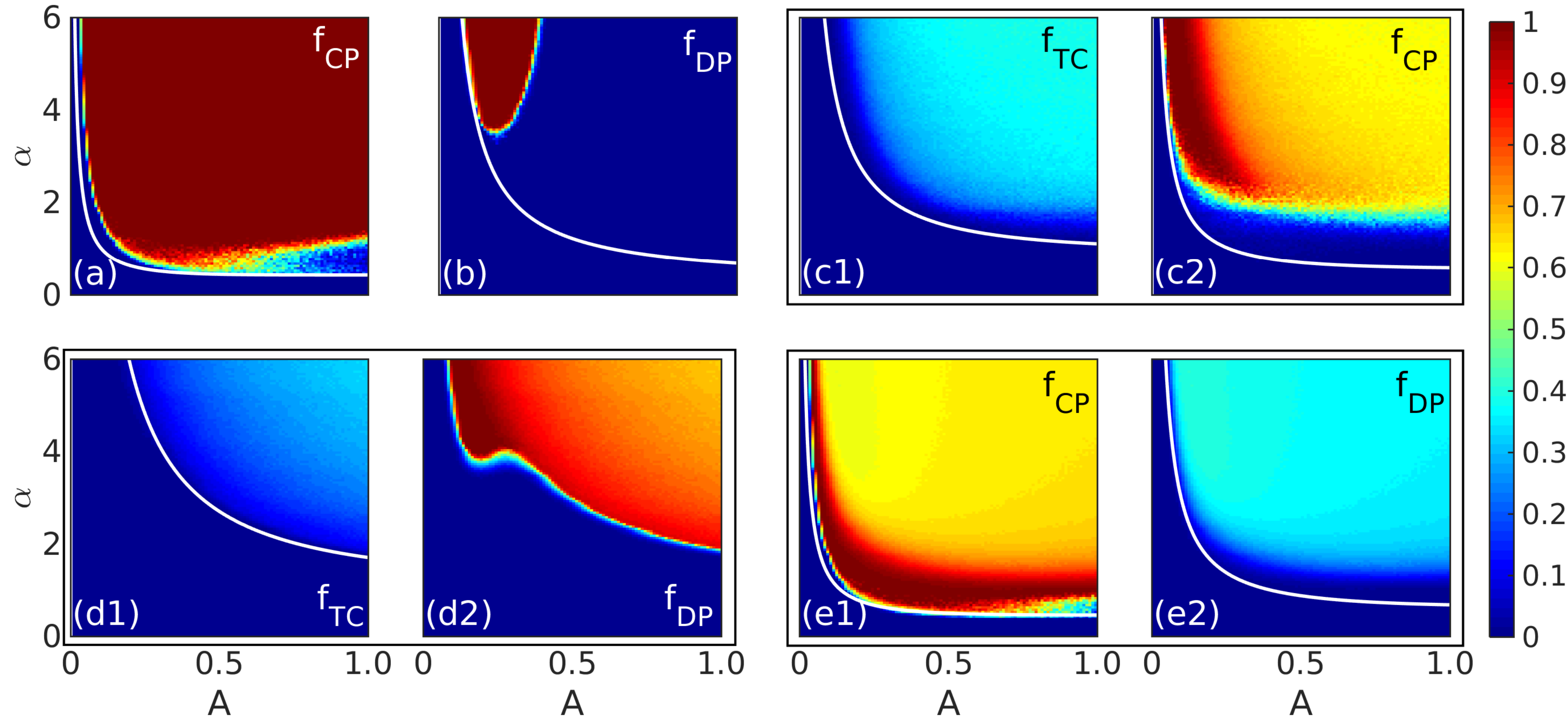}
\vspace{0cm}
\caption{The dependence of simulated final steady fractions of three nondefection strategies on both $A$ and $\alpha$, where the results obtained from networked populations are respectively illustrated for five different evolution situations: (a) D+CP, (b) D+DP, (c1) and (c2) TC+D+CP, (d1) and (d2) TC+D+DP, (e1) and (e2) D+CP+DP. In all cases, the value of synergy factor is $r=4.0$. Correspondingly, the value of $f_{s}$ used to semi-analytically estimate the boundary lines are (see Appendix~C for further details): (a) $f_{CP}=1.0$, (b) $f_{DP}=0.96$, (c1) $f_C=0.1$ and $f_{CP}=0.55$, (c2) $f_C=0.1$ and $f_{CP}=0.71$ (d1) $f_{TC}=0.115$ and $f_{DP}=0.465$, (e1) $f_{CP}=0.1$ and $f_{DP}=0.84$, (e2) $f_{CP}=0.3$ and $f_{DP}=0.45$.
}
\label{fig:newconditionnofour}
\end{figure*}

Using the same initial conditions for Fig.~\ref{fig:analysisnofour}, in Fig.~\ref{fig:newconditionnofour} we present the comprehensive picture in the parameter plane of $(A,~\alpha)$ of the evolutionary dynamics, as well as semi-analytically estimated boundary lines. Larger regions of ND phases suggest that network reciprocity induced by network structure largely benefits cooperation by allowing nondefectors to organize
themselves into compact clusters.  

\section*{Appendix C}
\label{sec:appendixc}
This appendix section presents how to semi-analytically obtain the boundaries separating SP and FD phase at which the following relationship
\begin{eqnarray} 
\Pi_{x}=\Pi_{D}
\label{eq:relationship}
\end{eqnarray} is satisfied. While $\Pi_{x}$ can be obtained according to the following method:
\begin{eqnarray}
\Pi_{X}=\sum\limits_{i}w_{s}\Pi_{s}.
\label{eq:payx}
\end{eqnarray} $w_{s}$ represents the contribution weight of population with strategy $s\in \{TC,~D,~CP,~DP\}$ in resisting defection. 
Furthermore, we can estimate the values of $f_s$ and $w_{s}$ based on either the initial values of $f_s$ for numerical integration of the equations given in Appendix~A for analytical treatments, or the proportions of different strategy populations near the boundaries between the two phases for simulations; so as to semi-analytically identify the boundary lines for corresponding evolution situations. In more detail, combing Eqs.~(\ref{eq:relationship})~(\ref{eq:payx}) with the theoretical expressions of payoffs for different strategies in Appendix~\ref{sec:appendixa}, we further accordingly give the  expressions of $\alpha$ at boundary lines as function of $A$ for the following six evolution situations: (1) D+CP:
\begin{widetext}
\begin{eqnarray}
\alpha & = & \frac{\frac{r}{G}-1}{E\psi +E(1.0-g^{'}_{CP})^{i}-1};\\ 
\quad E & = & \sum\limits_{i=0}^{G-1}\frac{(G-1)!}{i!(G-1-i)!}f^{i}_{CP}(1.0-f_{CP})^{G-1-i}, \quad \psi=\sum\limits^{i}_{j=0}\frac{i!}{j!(i-j)!}g^{j+1}_{CP}(1.0-g_{CP})^{i-j}\frac{G-1-i}{j+1}; \notag \\
\quad g^{'}_{CP} & = & \frac{Ai}{G} \quad g_{CP}=\frac{A(i+1)}{G}. \notag 
\label{eq:boundarydcp}
\end{eqnarray}
\end{widetext}
It should be noted that $E$ and $\psi$ are just two operators to make the equation look short, rather than functions or something else. Moreover, $f_{s}$ indicates the proportion of population of strategy $s$ at the boundary lines. 
(2) D+DP:
\begin{widetext}
\begin{eqnarray}
\alpha & = & \frac{\frac{r}{G}-1}{E\psi +E(1.0-g^{'}_{DP})^{i}-1};\\ 
\quad E & = & \sum\limits_{i=0}^{G-1}\frac{(G-1)!}{i!(G-1-i)!}f^{i}_{DP}(1.0-f_{DP})^{G-1-i}, \quad \psi=\sum\limits^{i}_{j=0}\frac{i!}{j!(i-j)!}g^{j+1}_{DP}(1.0-g_{DP})^{i-j}\frac{G-1-i}{j+1}; \notag \\
\quad g^{'}_{DP} & = & \frac{A(G-i)}{G} \quad g_{DP}=\frac{A(G-i-1)}{G}. \notag
\label{eq:boundaryddp} 
\end{eqnarray}
\end{widetext}
(3) TC+D+CP:
\begin{widetext}
\begin{eqnarray}
\alpha & = & \frac{\frac{r}{G}-1}{w_{CP}E\psi +E(1.0-g^{'}_{CP})^{i}-1}, \quad w_{CP}=\frac{f_{CP}}{f_{TC}+f_{CP}};\\ 
\quad E & = & \sum\limits_{i,j=0}^{G-1}\frac{(G-1)!}{i!j!(G-1-i-j!)!}f^{i}_{CP}f^{j}_{TC}(1.0-f_{CP}-f_{TC})^{G-1-i-j}, \quad \psi=\sum\limits^{i}_{k=0}\frac{i!}{k!(i-k)!}g^{k+1}_{CP}(1.0-g_{CP})^{i-k}\frac{G-1-i-j}{k+1}; \notag \\
\quad g^{'}_{CP} & = & \frac{A(i+j)}{G} \quad g_{CP}=\frac{A(i+j+1)}{G}. \notag 
\label{eq:boundarycdcp} 
\end{eqnarray}
\end{widetext}
(4) TC+D+DP:
\begin{widetext}
\begin{eqnarray}
\alpha & = & \frac{\frac{r}{G}-1}{w_{DP}E\psi +E(1.0-g^{'}_{DP})^{i}-1}, \quad w_{DP}=\frac{f_{DP}}{f_{TC}+f_{DP}};\\ 
\quad E & = & \sum\limits_{i,j=0}^{G-1}\frac{(G-1)!}{i!j!(G-1-i-j!)!}f^{i}_{DP}f^{j}_{TC}(1.0-f_{DP}-f_{TC})^{G-1-i-j}, \quad \psi=\sum\limits^{i}_{k=0}\frac{i!}{k!(i-k)!}g^{k+1}_{DP}(1.0-g_{DP})^{i-k}\frac{G-1-i-j}{k+1}; \notag \\
\quad g^{'}_{DP} & = & \frac{A(G-i-j)}{G} \quad g_{DP}=\frac{A(G-i-j-1)}{G}. \notag
\label{eq:boundarycddp}  
\end{eqnarray}
\end{widetext}
(5) D+CP+DP:
\begin{widetext}
\begin{eqnarray}
\alpha & = & \frac{\frac{r}{G}-1}{w_{DP}E\psi_{DP}+w_{CP}E\psi_{CP} +E(1.0-g^{'}_{CP})^{i}(1.0-g^{'}_{DP})^{j}-1};\\ 
w_{DP} & = & \frac{f_{DP}}{f_{CP}+f_{DP}}, \quad w_{CP}=\frac{f_{CP}}{f_{CP}+f_{DP}};\notag \\
E & = & \sum\limits_{i,j=0}^{G-1}\frac{(G-1)!}{i!j!(G-1-i-j!)!}f^{i}_{CP}f^{j}_{DP}(1.0-f_{CP}-f_{DP})^{G-1-i-j};\notag \\
\psi_{DP} & = & \sum\limits^{i}_{k=0}\frac{i!}{k!(i-k)!}g^{k}_{CP}(1.0-g_{CP})^{i-k}\sum\limits^{j}_{l=0}\frac{j!}{l!(j-l)!}g^{l+1}_{DP}(1.0-g_{DP})^{j-l}\frac{G-1-i-j}{k+l+1}; \notag \\
\psi_{CP} & = & \sum\limits^{i}_{k=0}\frac{i!}{k!(i-k)!}g^{k+1}_{CP}(1.0-g_{CP})^{i-k}\sum\limits^{j}_{l=0}\frac{j!}{l!(j-l)!}g^{l}_{DP}(1.0-g_{DP})^{j-l}\frac{G-1-i-j}{k+l+1}; \notag \\
g^{'}_{DP} & = & \frac{A(G-i-j)}{G},\quad g^{'}_{CP} = \frac{A(i+j)}{G}, \quad g_{DP}=\frac{A(G-i-j-1)}{G}, \quad g_{CP} = \frac{A(i+j+1)}{G}. \notag 
\label{eq:boundaryccpdp} 
\end{eqnarray}
\end{widetext}
(6) TC+D+CP+DP:
\begin{widetext}
\begin{eqnarray}
\alpha & = & \frac{\frac{r}{G}-1}{w_{CP}E\psi_{CP}+w_{DP}E\psi_{DP} +E(1.0-g^{'}_{CP})^{j}(1.0-g^{'}_{DP})^{k}-1};\\ 
w_{DP} & = & \frac{f_{DP}}{f_{TC}+f_{CP}+f_{DP}}, \quad w_{CP}=\frac{f_{CP}}{f_{TC}+f_{CP}+f_{DP}};\notag \\
E & = & \sum\limits_{i,j=0}^{G-1}\frac{(G-1)!}{i!j!k!(G-1-i-j-k)!}f^{i}_{TC}f^{j}_{CP}f^{k}_{DP}(1.0-f_{TC}-f_{CP}-f_{DP})^{G-1-i-j-k};\notag \\
\psi_{CP} & = & \sum\limits^{j}_{l=0}\frac{j!}{l!(j-l)!}g^{l+1}_{CP}(1.0-g_{CP})^{j-l}\sum\limits^{k}_{m=0}\frac{k!}{m!(k-m)!}g^{m}_{DP}(1.0-g_{DP})^{k-m}\frac{G-1-i-j-k}{l+m+1}; \notag \\
\psi_{DP} & = & \sum\limits^{j}_{l=0}\frac{j!}{l!(j-l)!}g^{l}_{CP}(1.0-g_{CP})^{j-l}\sum\limits^{k}_{m=0}\frac{k!}{m!(k-m)!}g^{m+1}_{DP}(1.0-g_{DP})^{k-m}\frac{G-1-i-j-k}{l+m+1}; \notag \\
g^{'}_{DP} & = & \frac{A(G-i-j-k)}{G},\quad g^{'}_{CP} = \frac{A(i+j+k)}{G}, \quad g_{DP}=\frac{A(G-i-j-k-1)}{G}, \quad g_{CP} = \frac{A(i+j+k+1)}{G}. \notag 
\label{eq:boundarycdcpdp}
\end{eqnarray}
\end{widetext}
However, Figs.~\ref{fig:analysisnofour}, \ref{fig:ratioanalysisfour}(d)-(e), \ref{fig:simulationnofour} and \ref{fig:simulationfour} show that it is very hard to get an accurate values of proportions different strategies i.~e., the values of $w_{s}$, because of high fluctuations near the boundaries between SP and FD phase, especially for simulation cases. Therefore, above equations actually provide a semi-analytical method to identify the boundary lines, because one have to estimate the values of $f_{s}$ so as to obtain a line which is close to the numerical boundaries as much as possible.  

\bibliography{multireference}
\end{document}